\documentclass[10pt,journal,compsoc]{IEEEtran}

\ifCLASSOPTIONcompsoc
  \usepackage[nocompress]{cite}
\else
  \usepackage{cite}
\fi

\usepackage{epsfig}
\usepackage{graphicx}

\usepackage{booktabs}
\usepackage[T1]{fontenc}
\newcommand{\changefont}[3]{
\fontfamily{#1} \fontseries{#2} \fontshape{#3} \selectfont}

\begin{document}
\title{STEVE -- Space-Time-Enclosing\\
Volume Extraction}

\author{B.~R.~Schlei
\IEEEcompsocitemizethanks{
\IEEEcompsocthanksitem B.~R.~Schlei is with the 
GSI Helmholtz Centre for Heavy Ion Research GmbH,
Planckstra\ss e 1, 64291 Darmstadt, Germany.}
}

\IEEEtitleabstractindextext{
\begin{abstract}
The novel STEVE (\textit{i.e.}, Space-Time-Enclosing Volume
Extraction) algorithm is described here for the very first time.
It generates iso-valued hypersurfaces that may be implicitly contained in
four-dimensional ($4$D) data sets, such as temporal sequences of
three-dimensional images from time-varying computed tomography.
Any final hypersurface that will be generated by STEVE is guaranteed to 
be free from accidental rifts, i.e., it always fully encloses a region
in the $4$D space under consideration.
Furthermore, the information of the interior/exterior of the enclosed
regions is propagated to each one of the tetrahedrons, which are embedded
into $4$D and which in their union represent the final, iso-valued
hypersurface(s). 
STEVE is usually executed in a purely data-driven mode, and it uses 
lesser computational resources than other techniques that also generate 
simplex-based manifolds of codimension $1$.
\end{abstract}
}

\maketitle

\IEEEpeerreviewmaketitle

\IEEEraisesectionheading{\section{Introduction}
\label{sec:introduction}
}

\IEEEPARstart{T}{he} following presentation is subject to a pending patent application
(\textit{cf.}, Ref.~\cite{PAT11}).
\\
\indent
Recent advances in the field of computed tomography (\textit{cf.}, e.g., 
Ref.s~\cite{XLI11}--\cite{SSL12}) have made available high-quality 
$4$D~\footnote{I.e., three spatial dimensions ($3$D) and one temporal 
dimension ($+1$D).} 
reconstructed sets of measured time-varying voxel data.
Similarly structured data may be generated while simulating $4$D 
transversal phase-spaces in ion beam transport, or in $3+1$D
fireball simulations in the field of theoretical heavy-ion physics,
just to name a few more examples (\textit{cf.}, 
section~\ref{sec:applications}).
Given such \textit{discrete} $4$D data, its further processing may 
require the extraction or generation of a certain number of
\textit{continuous}, iso-valued hypersurfaces that may or may not be 
contained in the data.
\\
\indent
Quite a number of papers deal with the subject of iso-(hyper)surface
construction in higher dimensions already (\textit{cf.}, e.g., 
Ref.s~\cite{FIDR96}--\cite{HUOV12}).
Their approaches are usually based on methods that can be used also for 
contour and surface extraction in two dimensions ($2$D) and $3$D, 
respectively.
E.g., in Ref.s~\cite{WEIG96,WEIG98}, the $4$D data are extruded to the
fifth dimension ($5$D) while representing them as a set of 
$4$-simplices~\footnote{I.e., pentachorons, pentatopes, or tetrahedral 
hyperpyramids.}.
Iso-hypersurfaces are then constructed while intersecting the 
$4$-simplices at a constant level in $5$D.
However, this technique does not ensure the construction of 
manifolds of codimension $1$. 
\\
\indent
Analogously, Weigle and Banks' method would represent a $2$D 
gray-level image as a triangular surface relief in $3$D.
In order to construct $1$D contours, the relief would be intersected 
with a $2$D plane at fixed iso-value.
However, since relief triangles could be oriented parallel to and totally
coincide with the intersecting plane, the resulting object would not
just simply consist of line segments, but -- perhaps -- also of triangles.
Note that in this paper we are especially concerned with the generation of
simplex-based manifolds of codimension $1$, \textit{i.e.}, hypersurfaces 
that \textit{purely} consist of $3$-simplices (tetrahedrons), which 
are embedded into the considered $4$D spaces.
\\
\indent
The latter requirement is fulfilled by the work of other authors, where
some of these algorithms could be categorized as follows:
those algorithms, which (i) use templates (\textit{cf.}, e.g.,
Ref.s~\cite{BHAN04,LACH00}), or
(ii) construct a mesh via (oriented) edge insertion, and which implicitly
solve ambiguities (\textit{cf.}, 
Ref.~\cite{HUOV12}~\footnote{Another algorithm 
is provided by the much earlier work of Fidrich~\cite{FIDR96}}), 
or (iii) are \textit{protomesh}-based, and which solve spatial and 
temporal ambiguities explicitly (this present work; \textit{cf.}, 
Ref.~\cite{BRS04} for a first announcement).
\\
\indent
This paper is organized as follows.
First we shall shortly revisit VESTA~\cite{BRS12}, i.e., an
algorithm that generates iso-surfaces in $3$D from volumetric 
(e.g., image) data.
Next, we shall present a common flowchart for the various processing 
stages of both VESTA and STEVE, because both algorithms work in a very
similar way.
While presenting the volume extraction framework, we are going to 
introduce the indexing scheme used by STEVE for the $4$D spaces under
consideration, as well as a complete, corresponding vector path 
table (\textit{cf.}, Table~\ref{tab:02}).
This table will provide all possible and/or necessary links for 
volume segments that contribute to the final hypersurfaces.
\\
\indent
Special emphasis will be put onto the consistent treatment of 
topological ambiguities.
In particular, we shall demonstrate that in \textit{discretized} 
$4$D spaces \textit{more than one} solution is generally possible 
for the constructed iso-hypersurfaces.
In an application related section, we provide a few examples for 
hypersurface extraction.
And last but not least, we discuss STEVE in comparison to other
$4$D hypersurface generation algorithms. 
This paper will conclude with a short summary.

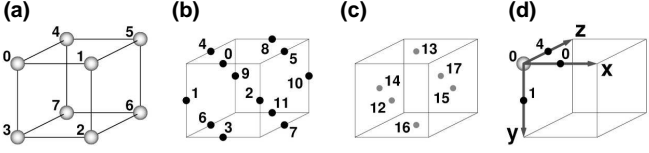
\begin{figure}[t]
  \begin{center}
    \scalebox{0.58}{
    \epsfig{width=15.25cm,figure=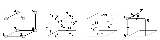}
    \setlength{\unitlength}{1cm}
    \put(-15.39,3.1){\changefont{phv}{b}{n}\Large (a)}
    \put(-11.53,3.1){\changefont{phv}{b}{n}\Large (b)}
    \put(-7.67,3.1){\changefont{phv}{b}{n}\Large (c)}
    \put(-3.81,3.1){\changefont{phv}{b}{n}\Large (d)}
    }
  \end{center}
  \vspace{-0.3cm} 
  \caption{
Indexing scheme for a $2\times2\times2$-neighborhood of voxels;
(a) voxel site IDs;
(b) boundary face centers;
(c) junctures;
(d) range vectors for voxel no. $0$, coinciding with the $3$D 
coordinate system $x|y|z$, and superimposed with corresponding 
boundary face centers.
} 
  \label{fig:01}
\end{figure}

\begin{table}[b]
\caption{
Directed paths for the quadrants of the oriented boundary faces
}
\label{tab:01}
\begin{center}
\begin{tabular}{rll}
\toprule
\textbf{Center ID$\:$}
&\boldmath$\oplus\:$\textbf{Path}
&\boldmath$\ominus\:$\textbf{Path}\\
\addlinespace[0.5ex]
\toprule
{\boldmath$0\:$}
&$\: 13 \rightarrow 0\rightarrow 12 \:$
&$\: 12 \rightarrow 0\rightarrow 13 \:$\\
{\boldmath$1\:$}
&$\: 12 \rightarrow 1\rightarrow 14 \:$
&$\: 14 \rightarrow 1\rightarrow 12 \:$\\
{\boldmath$2\:$}
&$\: 15 \rightarrow 2\rightarrow 12 \:$
&$\: 12 \rightarrow 2\rightarrow 15 \:$\\
{\boldmath$3\:$}
&$\: 12 \rightarrow 3\rightarrow 16 \:$
&$\: 16 \rightarrow 3\rightarrow 12 \:$\\
{\boldmath$4\:$}
&$\: 14 \rightarrow 4\rightarrow 13 \:$
&$\: 13 \rightarrow 4\rightarrow 14 \:$\\
{\boldmath$5\:$}
&$\: 13 \rightarrow 5\rightarrow 15 \:$
&$\: 15 \rightarrow 5\rightarrow 13 \:$\\
{\boldmath$6\:$}
&$\: 16 \rightarrow 6\rightarrow 14 \:$
&$\: 14 \rightarrow 6\rightarrow 16 \:$\\
{\boldmath$7\:$}
&$\: 15 \rightarrow 7\rightarrow 16 \:$
&$\: 16 \rightarrow 7\rightarrow 15 \:$\\
{\boldmath$8\:$}
&$\: 17 \rightarrow 8\rightarrow 13 \:$
&$\: 13 \rightarrow 8\rightarrow 17 \:$\\
{\boldmath$9\:$}
&$\: 14 \rightarrow 9\rightarrow 17 \:$
&$\: 17 \rightarrow 9\rightarrow 14 \:$\\
{\boldmath$10\:$}
&$\: 17 \rightarrow 10\rightarrow 15 \:$
&$\: 15 \rightarrow 10\rightarrow 17 \:$\\
{\boldmath$11\:$}
&$\: 16 \rightarrow 11\rightarrow 17 \:$
&$\: 17 \rightarrow 11\rightarrow 16 \:$\\
\bottomrule
\end{tabular}
\end{center}
\textbf{Note:} The quadrants of the oriented boundary faces have their 
centers at the predefined locations as pictured in Fig.~\ref{fig:01}.b. 
The start and end points of the $24$ paths are potential
points of ambiguity, as shown in Fig.~\ref{fig:01}.c.
For the orientations, \textit{cf.}, Appendix~\ref{app:appA}.
\end{table}

\section{Three Dimensions}

\begin{figure}[t]
  \begin{center}
    \scalebox{0.58}{
    \epsfig{width=13.5cm,figure=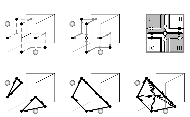} 
    \setlength{\unitlength}{1cm}
    \put(-13.73,7.53){\changefont{phv}{b}{n}\Large (a)}    
    \put(-9.105,7.53){\changefont{phv}{b}{n}\Large (b)}    
    \put(-4.48,7.53){\changefont{phv}{b}{n}\Large (c)}    
    \put(-13.73,3.33){\changefont{phv}{b}{n}\Large (d)}    
    \put(-9.105,3.33){\changefont{phv}{b}{n}\Large (e)}    
    \put(-4.48,3.33){\changefont{phv}{b}{n}\Large (f)}
    }    
  \end{center}
  \vspace{-0.3cm} 
  \caption{
Encounter of a topological ambiguity within a $3$D-cell:
(a) initial boundary faces;
(b) protomesh;
(c) connectivity diagram.
(d) two resulting oriented triangles from the ``disconnect'' mode;
(e) a final, more complex-shaped surface segment, resulting from 
the ``connect'' mode;
(f) as in (e), with a surface segment decomposisiton into six 
triangles.
} 
  \label{fig:02}
\end{figure}
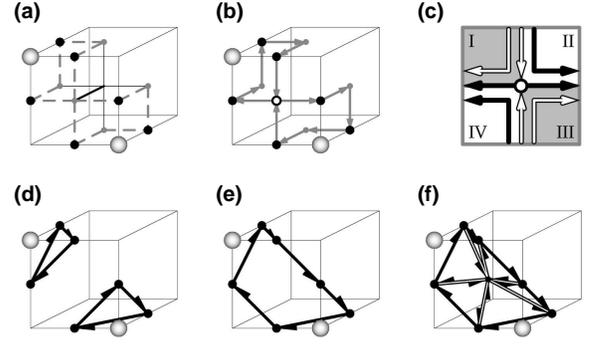

\begin{figure}[t]
  \begin{center}
    \scalebox{0.65}{
    \epsfig{width=7.0cm,figure=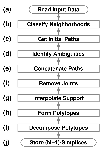} 
    }
  \end{center}
  \vspace{-0.3cm} 
  \caption{
Flowchart for both VESTA and STEVE (see text).
} 
  \label{fig:03}
\end{figure}

\noindent
The marching variant of VESTA~\cite{BRS12} treats each individual
$2\times2\times2$-neighborhood of a $3$D (e.g., image) data set 
separately.
In Fig.~\ref{fig:01}, we show the indexing scheme that VESTA uses
for each $3$D-cell, together with the $3$D coordinate sytem, $x|y|z$, 
as indicated in Fig.~\ref{fig:01}.d.
Each neighborhood cube has six faces that are each represented by the
junctures (gray dots) as shown in Fig.~\ref{fig:01}.c.
The cubes' twelve edges are represented by the boundary face centers
(black dots) as shown in Fig.~\ref{fig:01}.b,
whereas its eight corners are the centers of each voxel of the 
considered voxel neighborhood (\textit{cf.}, Fig.~\ref{fig:01}.a).
Fig.~\ref{fig:01}.d also shows the three range vectors of
voxel no. $0$, along which the final surface support 
points~\footnote{The boundary face centers are support points of the 
final surfaces (\textit{cf.}, Ref.~\cite{BRS12}, for more detail).} 
may be moved if interpolations are necessary.
\\
\indent
In addition to this indexing scheme, VESTA makes use of a
vector path table (\textit{cf.}, Table~\ref{tab:01}), which
provides all possible and/or necessary links for 
surface segments that contribute to the final iso-surfaces.
Next, we shall demonstrate the $3$D surface construction
with VESTA with the example shown in Fig.~\ref{fig:02} in conjunction
with the flowchart pictured in Fig.~\ref{fig:03}.
\\
\indent
After VESTA has read proper input data (\textit{cf.}, 
Fig.~\ref{fig:03}.a), a user-specified iso-value leads to
a segmentation, i.e., a classification of the 
$2\times2\times2$-neighborhoods 
(\textit{cf.}, Fig.~\ref{fig:03}.b).
In Fig.~\ref{fig:02}, we only the two voxels with the site IDs $0$ and $2$ 
have been selected while assuming that they have field values above or 
equal to a given iso-value (\textit{cf.}, Fig.~\ref{fig:02}, except
Fig.~\ref{fig:02}.c).
In the following, let us call selected voxels ``active'', and all 
remaining voxels ``inactive''.

\indent
VESTA introduces for each contact between an active and an inactive 
voxel a protomesh ``building block'' (PBB) as indicated in 
Fig.s~\ref{fig:05}.b, \ref{fig:06}.b, \&~\ref{fig:06}.c.
In Fig.~\ref{fig:02}.a, these show up as six quadrants, whereas
in Fig.~\ref{fig:02}.b, they are indicated by twelve voxel
face vectors (VFV) and six corresponding boundary face centers
(black dots) with the IDs $0$, $1$, $2$, $3$, $4$, and $7$.
In addition, Fig.~\ref{fig:02}.b shows five junctures (gray 
dots) with IDs $12$, $13$, $14$, $15$, and $16$.
The initial paths (\textit{cf.}, Fig.~\ref{fig:03}.c) of this cell
can be read off from Table~\ref{tab:01}.
These are the paths with no.s (center ID / orientation) $0 \:\oplus$, 
$1\:\oplus$, $2 \:\ominus$, $3 \:\ominus$, $4 \:\oplus$, and 
$7 \:\oplus$.
\\
\indent
The juncture with ID $12$ is a so-called ``point of
ambiguity'' (POA, \textit{cf.}, Fig.~\ref{fig:03}.d).
Where all other selected junctures feature exactly one incoming and
one outgoing VFV, the POA has two incoming and two outgoing VFVs.
The connectivity diagramm as shown in Fig.~\ref{fig:02}.c helps one
to resolve the consistent concatenation of VFVs (\textit{cf.},
Fig.~\ref{fig:03}.e).
If an incoming VFV that is part of the PBB of a given voxel will be 
connected to an outgoing VFV of another PBB of the \textit{same} 
voxel, then one operates in the so-called ``disconnect'' mode.
Otherwise, if one connects it to an outgoing VFV of another PBB of 
a \textit{different} voxel, then one operates in the so-called 
``connect'' mode.
\\
\indent
Either choice of concatenation leads to the formation of a $3$D 
protomesh (\textit{cf.}, Fig.~\ref{fig:02}.b).
In the ``disconnect'' mode one obtains here the two meshes
$13 \rightarrow 0 \rightarrow 12 \rightarrow 1 
\rightarrow 14 \rightarrow 4 \rightarrow 13$, and
$12 \rightarrow 2 \rightarrow 15 \rightarrow 7 
\rightarrow 16 \rightarrow 3 \rightarrow 12$, whereas
in the ``connect'' mode one obtains here the single mesh
$13 \rightarrow 0 \rightarrow 12 \rightarrow 2 
\rightarrow 15 \rightarrow 7 \rightarrow 16
\rightarrow 3 \rightarrow 12 \rightarrow 1 
\rightarrow 14 \rightarrow 4 \rightarrow 13$.
After the removal of the junctures (or joints, \textit{cf.},
Fig.~\ref{fig:03}.f), one is left with 
polytopes~\footnote{The constructed polytopes are shaped irregularly 
in general (\textit{cf.}, Ref.~\cite{COXE73}).} 
in $3$D (\textit{cf.}, Fig.~\ref{fig:03}.h).
\\
\indent
In the ``disconnect'' mode one obtains here the two vector cycles
$0  \rightarrow 1 \rightarrow 4 \rightarrow 0$, and
$2 \rightarrow 7 \rightarrow 3 \rightarrow 2$, as
shown in Fig.~\ref{fig:02}.d, whereas
in the ``connect'' mode one obtains here the single vector cycle
$0 \rightarrow 2 \rightarrow 7 \rightarrow 3 \rightarrow 1 
\rightarrow 4 \rightarrow 0$, as shown in Fig.~\ref{fig:02}.e.
Hence, in the ``disconnect'' mode we end up with two oriented
triangles (VESTA $3$-cycles), and no further processing is
necessary.
However, in the ``connect'' mode we end up with one oriented
VESTA $6$-cycle.
In Fig.~\ref{fig:02}.f, this vector cycle has been decomposed into 
six oriented triangles (\textit{cf.}, Fig.s~\ref{fig:03}.i 
and~\ref{fig:14}.6.b).
VESTA has stored at this stage the generated $2$-simplices 
(\textit{cf.}, Fig.~\ref{fig:03}.j) and thereby completes its 
processing.
Note that we did not demonstrate the interpolation of support 
points (\textit{cf.}, Fig.~\ref{fig:03}.g) in this example.

\section{The Volume Extraction Framework}
\label{sec:framework}

\noindent
Note that because of Ref.~\cite{PAT11}, we exhibit in this paper 
neither any source code nor any pseudocode of the STEVE algorithm.
However, we present -- among many other things -- a flowchart that
represents its various processing steps (\textit{cf.}, 
Fig.~\ref{fig:03}).
We shall address each single stage of this flowchart in one or more 
of the following subsections.

\begin{figure}[b]
  \begin{center}
    \scalebox{0.58}{
    \epsfig{width=12.0cm,figure=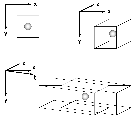} 
    \setlength{\unitlength}{1cm}
    \put(-12.31,6.42){\changefont{phv}{b}{n}\Large (a)}    
    \put(-6.99,6.42){\changefont{phv}{b}{n}\Large (b)}    
    \put(-12.31,0.44){\changefont{phv}{b}{n}\Large (c)}  
    }  
  \end{center}
  \vspace{-0.3cm} 
  \caption{
Coordinate systems in 
(a) $2$D with a single pixel;
(b) $3$D with a single voxel; and
(c) $4$D with a single toxel.
} 
  \label{fig:04}
\end{figure}

\subsection{Initial Mesh Features}

\noindent
The geometrical structure of the underlying meshes (or grids), which 
support the data that should be processed, is as follows.
Here, we shall consider purely homogeneous, $N$-cubical (cartesian) 
grids only, with $N$ referring to the integral dimension of the space 
under consideration.
We choose $N$, because we intend to compare our $4$D method with
analogue $2$D and $3$D techniques.
In particular, we shall use two different approaches.
Either we shall look at the data globally, or we shall look at a
$2(\times2)^{N-1}$-neighborhood of the data.
In the first case we shall use the term ``global view'' (GV), and in 
the last case we shall use the term ``neighborhood view'' (NV), 
respectively.
Furthermore, let the fourth dimension -- without loss of 
generality -- be named ``time''~\footnote{Note however, that we do 
not necessarily consider a metric here.}.
\\
\indent
In Fig.~\ref{fig:04}, various coordinate systems that we are going to use
($x|y$ in $2$D, $x|y|z$ in $3$D, and $x|y|z|t$ in $4$D, respectively)
are shown together with samples of single, corresponding picture elements 
(in GV).
The centers of the elements, i.e., a pixel, a voxel, and a toxel 
(i.e., time-varying voxel) are each marked with a sphere.
\\
\indent
The STEVE algorithm starts with the reading of proper input data
(\textit{cf.}, Fig.~\ref{fig:03}.a). 
This input should specify the overall dimensions of the $4$D space
under consideration, e.g., by indicating the bounds and maximum numbers 
of toxels of the homogeneous $4$D grid in the $x|y|z|t$-directions.
Furthermore, it should consist of the individual $x|y|z|t$-positions 
and one (or more~\footnote{This, of course, depends on the particular
application one may have in mind.}) field value(s) that is (are) 
associated with each center of the corresponding toxels.
\\
\indent
Next, we shall discuss the segmentation of the
data, i.e., the differentiation of picture elements (or data cells) 
that should be enclosed with a manifold of codimension $1$ from those 
that should not be enclosed.

\begin{figure}[t]
  \begin{center}
    \scalebox{0.58}{
    \epsfig{width=14.0cm,figure=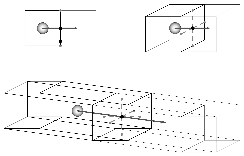} 
    \setlength{\unitlength}{1cm}
    \put(-14.29,5.72){\changefont{phv}{b}{n}\Large (a)}    
    \put(-7.50,5.72){\changefont{phv}{b}{n}\Large (b)}    
    \put(-14.29,0.50){\changefont{phv}{b}{n}\Large (c)} 
    }   
  \end{center}
  \vspace{-0.3cm} 
  \caption{
Pairs of picture elements that are in direct contact in 
(a) $2$D;
(b) $3$D; and 
(c) $4$D (see text).
} 
  \label{fig:05}
\end{figure}

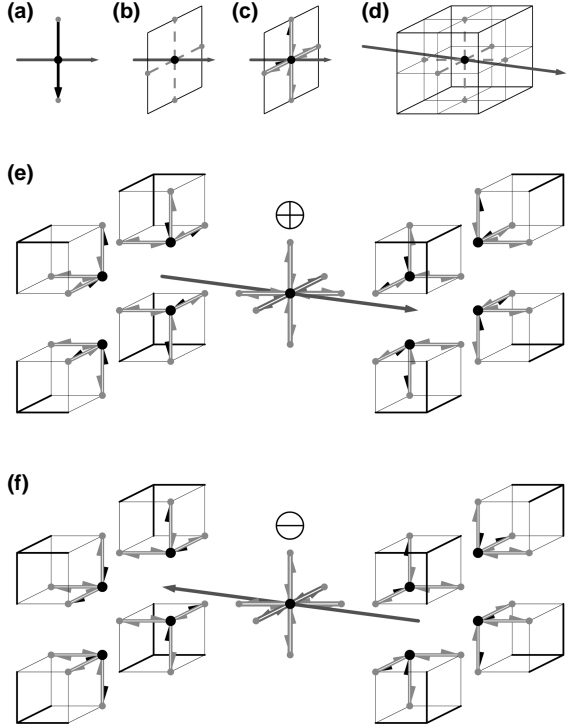
\begin{figure}[t]
  \begin{center}
    \scalebox{0.58}{
    \epsfig{width=13.0cm,figure=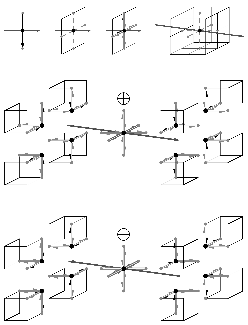} 
    \setlength{\unitlength}{1cm}
    \put(-13.37,16.39){\changefont{phv}{b}{n}\Large (a)}    
    \put(-10.95,16.39){\changefont{phv}{b}{n}\Large (b)}    
    \put(-8.22,16.39){\changefont{phv}{b}{n}\Large (c)}    
    \put(-5.25,16.39){\changefont{phv}{b}{n}\Large (d)}    
    \put(-13.37,12.68){\changefont{phv}{b}{n}\Large (e)}    
    \put(-13.37,5.56){\changefont{phv}{b}{n}\Large (f)}   
    } 
  \end{center}
  \vspace{-0.3cm} 
  \caption{
Initial protomesh building blocks for contour, surface and hypersurface
construction in 
(a) for $2$D, in 
(b) \& (c) for $3$D, and in
(d) -- (f) for $4$D, respectively (see text).
} 
  \label{fig:06}
\end{figure}

\subsection{Data Segmentation}

\noindent
After STEVE has obtained proper input, the specification of an 
iso-value will allow for the segmentation of these data.
I.e., toxels that have a (selected) field value larger than or 
equal to the iso-value will be marked for enclosure, whereas the 
others will not, or vice versa.
Hence, the segmentation of a $4$D space leads to a classification 
of its various toxel neighborhoods (\textit{cf.}, 
Fig.~\ref{fig:03}.b).
In Fig.~\ref{fig:05}, we show -- in GV -- pairs of picture elements 
in each $2$D, $3$D, and $4$D that are in direct contact.
In each dimension one element is active, i.e., it is marked for
enclosure (indicated by the spheres), and the other element is 
considered inactive, i.e., it should not be enclosed (no spheres 
are placed in their centers).
\\
\indent
In Fig.~\ref{fig:05}, each center of an active element is the origin 
of a range vector, which ends in the center of the inactive 
neighboring picture element.
These range vectors (dark gray; \textit{cf.}, Fig.~\ref{fig:06}) define 
the bounds within which a support point (black dots) of the final extracted, 
iso-valued manifolds may be repositioned.
E.g., the STEVE algorithm uses linear interpolation~\footnote{More 
complicated types of interpolations than just simple linear 
interpolations are possible.
E.g., one could use (higher-dimensional) B-Splines\cite{BOEH07,SALO05} 
instead, etc.} for this purpose at some later stage in the 
processing (\textit{cf.}, Fig.~\ref{fig:03}.g).
\\
\indent
The transitions from an active picture element to an inactive
one are of particular interest here.
They are shown once again in Fig.~\ref{fig:06}.
For the sake of completeness, Fig.~\ref{fig:06}.a depicts the 
building block that allows for contour extraction in $2$D with
DICONEX (\textit{cf.}, Ref.\cite{BRS09} for more
detail).
In $3$D, VESTA concatenates single squares with the help of VFVs, which
have been described in the previous section
(\textit{cf.}, Fig.s~\ref{fig:06}.b and~\ref{fig:06}.c).
In analogy, STEVE concatenates single cubes in $4$D.
The new $4$D building block, which aids this concatenation, is shown 
(in GV) in Fig.~\ref{fig:06}.d (dashed 
lines~\footnote{Dashed lines represent internal vector paths, 
so-called ``toxel cube vectors''.}); 
in Fig.~\ref{fig:06}.e, as exploded view for \textit{positive} 
orientation ($\oplus$, i.e., the range vector points into positive $x$-, 
$y$-, $z$-, or $t$-direction, respectively); 
in Fig.~\ref{fig:06}.f, as exploded view for \textit{negative} 
orientation ($\ominus$, i.e., the range vector points into negative $x$-, 
$y$-, $z$-, or $t$-direction, respectively).
\\
\indent
Note that in Fig.s~\ref{fig:06}.e and~\ref{fig:06}.f, the tiny black 
arrows mark toxel cube vectors (TCV) pairs, which belong together.
All TCV pairs connect a juncture of volumes (i.e., cubes) that are in 
contact to another volume through a $4$D support point (black dots) with 
another juncture (gray dots).

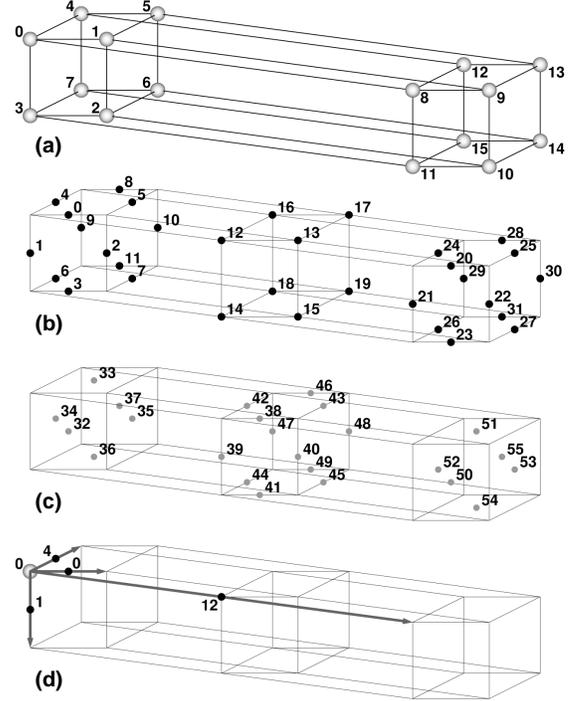
\begin{figure}[t]
  \begin{center}
    \scalebox{0.58}{
    \epsfig{width=13.0cm,figure=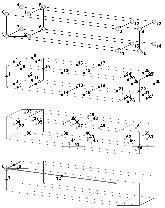} 
    \setlength{\unitlength}{1cm}
    \put(-12.74,12.72){\changefont{phv}{b}{n}\Large (a)}    
    \put(-12.74,8,64){\changefont{phv}{b}{n}\Large (b)}    
    \put(-12.74,4,56){\changefont{phv}{b}{n}\Large (c)}    
    \put(-12.74,0.48){\changefont{phv}{b}{n}\Large (d)}  
    }  
  \end{center}
  \vspace{-0.3cm} 
  \caption{
Indexing scheme for a $2\times2\times2\times2$-neighborhood of
toxels;
(a) toxel site IDs;
(b) boundary cube centers;
(c) junctures;
(d) range vectors for toxel no. $0$, superimposed with
corresponding boundary cube centers.
} 
  \label{fig:07}
\end{figure}

\subsection{Indexing Scheme and Vector Paths}

\noindent
It is sufficient to discuss all transitions from active to inactive toxels 
within single $2\times2\times2\times2$-neighborhoods (or $4$D-cells).
Therefore, we label each relevant $4$D point of a given $4$D-cell with a 
unique index.
\\
\indent
In Fig.~\ref{fig:07}, we show the notations that we have chosen for a given 
$4$D-cell (in NV).
In Fig.~\ref{fig:07}.a, all sixteen toxel site IDs are shown.
Since we have named the fourth dimension time (\textit{cf.}, above),
let No.s $0$ through $7$ represent the ``past,'' and no.s $8$ 
through $15$ the ``future''.
In Fig.~\ref{fig:07}.b, all thirty-two potential boundary cube centers 
(i.e., final manifold support points) are pictured.
And finally, in Fig.~\ref{fig:07}.c, all twenty-four junctures
-- which are potential points of ambiguity (POA) -- are shown for 
neighboring boundary cubes.
As an example, Fig.~\ref{fig:07}.d, shows for toxel no.~$0$ four range 
vectors (with positive (i.e., $\oplus$) orientations each in the $x$-,  
$y$-,  $z$-, and $t$-directions), superimposed with their 
corresponding boundary cube centers.
Note that the central cubes in Fig.s~\ref{fig:07}.b--d represent in
this chosen context the ``present''.

\begin{table*}[t]
\caption{
Triplets of directed paths for the octants of the oriented boundary cubes
}
\label{tab:02}
\begin{center}
\begin{tabular}{rllllll}
\toprule
\textbf{Center ID$\:$}
&\boldmath$\oplus\:$\textbf{Path 1}
&\boldmath$\oplus\:$\textbf{Path 2}
&\boldmath$\oplus\:$\textbf{Path 3}
&\boldmath$\ominus\:$\textbf{Path 1}
&\boldmath$\ominus\:$\textbf{Path 2}
&\boldmath$\ominus\:$\textbf{Path 3}\\
\addlinespace[0.5ex]
\toprule
{\boldmath$0\:$}
&$\: 33 \rightarrow 0\rightarrow 32 \:$
&$\: 32 \rightarrow 0\rightarrow 38 \:$
&$\: 38 \rightarrow 0\rightarrow 33 \:$
&$\: 32 \rightarrow 0\rightarrow 33 \:$
&$\: 33 \rightarrow 0\rightarrow 38 \:$
&$\: 38 \rightarrow 0\rightarrow 32 \:$\\
{\boldmath$1\:$}
&$\: 32 \rightarrow 1\rightarrow 34 \:$
&$\: 34 \rightarrow 1\rightarrow 39 \:$
&$\: 39 \rightarrow 1\rightarrow 32 \:$
&$\: 34 \rightarrow 1\rightarrow 32 \:$
&$\: 32 \rightarrow 1\rightarrow 39 \:$
&$\: 39 \rightarrow 1\rightarrow 34 \:$\\
{\boldmath$2\:$}
&$\: 35 \rightarrow 2\rightarrow 32 \:$
&$\: 32 \rightarrow 2\rightarrow 40 \:$
&$\: 40 \rightarrow 2\rightarrow 35 \:$
&$\: 32 \rightarrow 2\rightarrow 35 \:$
&$\: 35 \rightarrow 2\rightarrow 40 \:$
&$\: 40 \rightarrow 2\rightarrow 32 \:$\\
{\boldmath$3\:$}
&$\: 32 \rightarrow 3\rightarrow 36 \:$
&$\: 36 \rightarrow 3\rightarrow 41 \:$
&$\: 41 \rightarrow 3\rightarrow 32 \:$
&$\: 36 \rightarrow 3\rightarrow 32 \:$
&$\: 32 \rightarrow 3\rightarrow 41 \:$
&$\: 41 \rightarrow 3\rightarrow 36 \:$\\
{\boldmath$4\:$}
&$\: 34 \rightarrow 4\rightarrow 33 \:$
&$\: 33 \rightarrow 4\rightarrow 42 \:$
&$\: 42 \rightarrow 4\rightarrow 34 \:$
&$\: 33 \rightarrow 4\rightarrow 34 \:$
&$\: 34 \rightarrow 4\rightarrow 42 \:$
&$\: 42 \rightarrow 4\rightarrow 33 \:$\\
{\boldmath$5\:$}
&$\: 33 \rightarrow 5\rightarrow 35 \:$
&$\: 35 \rightarrow 5\rightarrow 43 \:$
&$\: 43 \rightarrow 5\rightarrow 33 \:$
&$\: 35 \rightarrow 5\rightarrow 33 \:$
&$\: 33 \rightarrow 5\rightarrow 43 \:$
&$\: 43 \rightarrow 5\rightarrow 35 \:$\\
{\boldmath$6\:$}
&$\: 36 \rightarrow 6\rightarrow 34 \:$
&$\: 34 \rightarrow 6\rightarrow 44 \:$
&$\: 44 \rightarrow 6\rightarrow 36 \:$
&$\: 34 \rightarrow 6\rightarrow 36 \:$
&$\: 36 \rightarrow 6\rightarrow 44 \:$
&$\: 44 \rightarrow 6\rightarrow 34 \:$\\
{\boldmath$7\:$}
&$\: 35 \rightarrow 7\rightarrow 36 \:$
&$\: 36 \rightarrow 7\rightarrow 45 \:$
&$\: 45 \rightarrow 7\rightarrow 35 \:$
&$\: 36 \rightarrow 7\rightarrow 35 \:$
&$\: 35 \rightarrow 7\rightarrow 45 \:$
&$\: 45 \rightarrow 7\rightarrow 36 \:$\\
{\boldmath$8\:$}
&$\: 37 \rightarrow 8\rightarrow 33 \:$
&$\: 33 \rightarrow 8\rightarrow 46 \:$
&$\: 46 \rightarrow 8\rightarrow 37 \:$
&$\: 33 \rightarrow 8\rightarrow 37 \:$
&$\: 37 \rightarrow 8\rightarrow 46 \:$
&$\: 46 \rightarrow 8\rightarrow 33 \:$\\
{\boldmath$9\:$}
&$\: 34 \rightarrow 9\rightarrow 37 \:$
&$\: 37 \rightarrow 9\rightarrow 47 \:$
&$\: 47 \rightarrow 9\rightarrow 34 \:$
&$\: 37 \rightarrow 9\rightarrow 34 \:$
&$\: 34 \rightarrow 9\rightarrow 47 \:$
&$\: 47 \rightarrow 9\rightarrow 37 \:$\\
{\boldmath$10\:$}
&$\: 37 \rightarrow 10\rightarrow 35 \:$
&$\: 35 \rightarrow 10\rightarrow 48 \:$
&$\: 48 \rightarrow 10\rightarrow 37 \:$
&$\: 35 \rightarrow 10\rightarrow 37 \:$
&$\: 37 \rightarrow 10\rightarrow 48 \:$
&$\: 48 \rightarrow 10\rightarrow 35 \:$\\
{\boldmath$11\:$}
&$\: 36 \rightarrow 11\rightarrow 37 \:$
&$\: 37 \rightarrow 11\rightarrow 49 \:$
&$\: 49 \rightarrow 11\rightarrow 36 \:$
&$\: 37 \rightarrow 11\rightarrow 36 \:$
&$\: 36 \rightarrow 11\rightarrow 49 \:$
&$\: 49 \rightarrow 11\rightarrow 37 \:$\\
{\boldmath$12\:$}
&$\: 38 \rightarrow 12\rightarrow 39 \:$
&$\: 39 \rightarrow 12\rightarrow 42 \:$
&$\: 42 \rightarrow 12\rightarrow 38 \:$
&$\: 39 \rightarrow 12\rightarrow 38 \:$
&$\: 38 \rightarrow 12\rightarrow 42 \:$
&$\: 42 \rightarrow 12\rightarrow 39 \:$\\
{\boldmath$13\:$}
&$\: 40 \rightarrow 13\rightarrow 38 \:$
&$\: 38 \rightarrow 13\rightarrow 43 \:$
&$\: 43 \rightarrow 13\rightarrow 40 \:$
&$\: 38 \rightarrow 13\rightarrow 40 \:$
&$\: 40 \rightarrow 13\rightarrow 43 \:$
&$\: 43 \rightarrow 13\rightarrow 38 \:$\\
{\boldmath$14\:$}
&$\: 39 \rightarrow 14\rightarrow 41 \:$
&$\: 41 \rightarrow 14\rightarrow 44 \:$
&$\: 44 \rightarrow 14\rightarrow 39 \:$
&$\: 41 \rightarrow 14\rightarrow 39 \:$
&$\: 39 \rightarrow 14\rightarrow 44 \:$
&$\: 44 \rightarrow 14\rightarrow 41 \:$\\
{\boldmath$15\:$}
&$\: 41 \rightarrow 15\rightarrow 40 \:$
&$\: 40 \rightarrow 15\rightarrow 45 \:$
&$\: 45 \rightarrow 15\rightarrow 41 \:$
&$\: 40 \rightarrow 15\rightarrow 41 \:$
&$\: 41 \rightarrow 15\rightarrow 45 \:$
&$\: 45 \rightarrow 15\rightarrow 40 \:$\\
{\boldmath$16\:$}
&$\: 47 \rightarrow 16\rightarrow 46 \:$
&$\: 46 \rightarrow 16\rightarrow 42 \:$
&$\: 42 \rightarrow 16\rightarrow 47 \:$
&$\: 46 \rightarrow 16\rightarrow 47 \:$
&$\: 47 \rightarrow 16\rightarrow 42 \:$
&$\: 42 \rightarrow 16\rightarrow 46 \:$\\
{\boldmath$17\:$}
&$\: 46 \rightarrow 17\rightarrow 48 \:$
&$\: 48 \rightarrow 17\rightarrow 43 \:$
&$\: 43 \rightarrow 17\rightarrow 46 \:$
&$\: 48 \rightarrow 17\rightarrow 46 \:$
&$\: 46 \rightarrow 17\rightarrow 43 \:$
&$\: 43 \rightarrow 17\rightarrow 48 \:$\\
{\boldmath$18\:$}
&$\: 49 \rightarrow 18\rightarrow 47 \:$
&$\: 47 \rightarrow 18\rightarrow 44 \:$
&$\: 44 \rightarrow 18\rightarrow 49 \:$
&$\: 47 \rightarrow 18\rightarrow 49 \:$
&$\: 49 \rightarrow 18\rightarrow 44 \:$
&$\: 44 \rightarrow 18\rightarrow 47 \:$\\
{\boldmath$19\:$}
&$\: 48 \rightarrow 19\rightarrow 49 \:$
&$\: 49 \rightarrow 19\rightarrow 45 \:$
&$\: 45 \rightarrow 19\rightarrow 48 \:$
&$\: 49 \rightarrow 19\rightarrow 48 \:$
&$\: 48 \rightarrow 19\rightarrow 45 \:$
&$\: 45 \rightarrow 19\rightarrow 49 \:$\\
{\boldmath$20\:$}
&$\: 50 \rightarrow 20\rightarrow 51 \:$
&$\: 51 \rightarrow 20\rightarrow 38 \:$
&$\: 38 \rightarrow 20\rightarrow 50 \:$
&$\: 51 \rightarrow 20\rightarrow 50 \:$
&$\: 50 \rightarrow 20\rightarrow 38 \:$
&$\: 38 \rightarrow 20\rightarrow 51 \:$\\
{\boldmath$21\:$}
&$\: 52 \rightarrow 21\rightarrow 50 \:$
&$\: 50 \rightarrow 21\rightarrow 39 \:$
&$\: 39 \rightarrow 21\rightarrow 52 \:$
&$\: 50 \rightarrow 21\rightarrow 52 \:$
&$\: 52 \rightarrow 21\rightarrow 39 \:$
&$\: 39 \rightarrow 21\rightarrow 50 \:$\\
{\boldmath$22\:$}
&$\: 50 \rightarrow 22\rightarrow 53 \:$
&$\: 53 \rightarrow 22\rightarrow 40 \:$
&$\: 40 \rightarrow 22\rightarrow 50 \:$
&$\: 53 \rightarrow 22\rightarrow 50 \:$
&$\: 50 \rightarrow 22\rightarrow 40 \:$
&$\: 40 \rightarrow 22\rightarrow 53 \:$\\
{\boldmath$23\:$}
&$\: 54 \rightarrow 23\rightarrow 50 \:$
&$\: 50 \rightarrow 23\rightarrow 41 \:$
&$\: 41 \rightarrow 23\rightarrow 54 \:$
&$\: 50 \rightarrow 23\rightarrow 54 \:$
&$\: 54 \rightarrow 23\rightarrow 41 \:$
&$\: 41 \rightarrow 23\rightarrow 50 \:$\\
{\boldmath$24\:$}
&$\: 51 \rightarrow 24\rightarrow 52 \:$
&$\: 52 \rightarrow 24\rightarrow 42 \:$
&$\: 42 \rightarrow 24\rightarrow 51 \:$
&$\: 52 \rightarrow 24\rightarrow 51 \:$
&$\: 51 \rightarrow 24\rightarrow 42 \:$
&$\: 42 \rightarrow 24\rightarrow 52 \:$\\
{\boldmath$25\:$}
&$\: 53 \rightarrow 25\rightarrow 51 \:$
&$\: 51 \rightarrow 25\rightarrow 43 \:$
&$\: 43 \rightarrow 25\rightarrow 53 \:$
&$\: 51 \rightarrow 25\rightarrow 53 \:$
&$\: 53 \rightarrow 25\rightarrow 43 \:$
&$\: 43 \rightarrow 25\rightarrow 51 \:$\\
{\boldmath$26\:$}
&$\: 52 \rightarrow 26\rightarrow 54 \:$
&$\: 54 \rightarrow 26\rightarrow 44 \:$
&$\: 44 \rightarrow 26\rightarrow 52 \:$
&$\: 54 \rightarrow 26\rightarrow 52 \:$
&$\: 52 \rightarrow 26\rightarrow 44 \:$
&$\: 44 \rightarrow 26\rightarrow 54 \:$\\
{\boldmath$27\:$}
&$\: 54 \rightarrow 27\rightarrow 53 \:$
&$\: 53 \rightarrow 27\rightarrow 45 \:$
&$\: 45 \rightarrow 27\rightarrow 54 \:$
&$\: 53 \rightarrow 27\rightarrow 54 \:$
&$\: 54 \rightarrow 27\rightarrow 45 \:$
&$\: 45 \rightarrow 27\rightarrow 53 \:$\\
{\boldmath$28\:$}
&$\: 51 \rightarrow 28\rightarrow 55 \:$
&$\: 55 \rightarrow 28\rightarrow 46 \:$
&$\: 46 \rightarrow 28\rightarrow 51 \:$
&$\: 55 \rightarrow 28\rightarrow 51 \:$
&$\: 51 \rightarrow 28\rightarrow 46 \:$
&$\: 46 \rightarrow 28\rightarrow 55 \:$\\
{\boldmath$29\:$}
&$\: 55 \rightarrow 29\rightarrow 52 \:$
&$\: 52 \rightarrow 29\rightarrow 47 \:$
&$\: 47 \rightarrow 29\rightarrow 55 \:$
&$\: 52 \rightarrow 29\rightarrow 55 \:$
&$\: 55 \rightarrow 29\rightarrow 47 \:$
&$\: 47 \rightarrow 29\rightarrow 52 \:$\\
{\boldmath$30\:$}
&$\: 53 \rightarrow 30\rightarrow 55 \:$
&$\: 55 \rightarrow 30\rightarrow 48 \:$
&$\: 48 \rightarrow 30\rightarrow 53 \:$
&$\: 55 \rightarrow 30\rightarrow 53 \:$
&$\: 53 \rightarrow 30\rightarrow 48 \:$
&$\: 48 \rightarrow 30\rightarrow 55 \:$\\
{\boldmath$31\:$}
&$\: 55 \rightarrow 31\rightarrow 54 \:$
&$\: 54 \rightarrow 31\rightarrow 49 \:$
&$\: 49 \rightarrow 31\rightarrow 55 \:$
&$\: 54 \rightarrow 31\rightarrow 55 \:$
&$\: 55 \rightarrow 31\rightarrow 49 \:$
&$\: 49 \rightarrow 31\rightarrow 54 \:$\\
\bottomrule
\end{tabular}
\end{center}
\textbf{Note:} The oriented boundary cubes have their centers at the 
predefined locations as pictured in Fig.~\ref{fig:07}.b. 
The start and end points of the $192$ paths are potential
points of ambiguity, as shown in Fig.~\ref{fig:07}.c.
For the orientations, \textit{cf.}, Appendix~\ref{app:appA}.
\end{table*}

\indent
As Fig.~\ref{fig:07}.d shows, each of the sixteen toxels 
may contribute within the $4$D-cell with exactly four boundary volume 
octants (i.e., an eights of a full boundary cube).
Note that each boundary volume octant (BVO) is represented 
within STEVE by a \textit{triplet} of TCV pairs (\textit{cf.}, 
Fig.s~\ref{fig:06}.e and~\ref{fig:06}.f).
Hence, one has in total $16\times4\times3=192$ different possible 
vector paths (i.e., TCV pairs) within a given toxel neighborhood.
The indexing scheme as shown in Fig.~\ref{fig:07} and the table of 
the $192$ vector paths (\textit{cf.}, Table~\ref{tab:02}) form two
of the three main ingredients that STEVE uses for iso-hypersurface 
construction.
\\
\indent
The third main ingredient for STEVE is a connectivity diagram 
(\textit{cf.}, Fig.~\ref{fig:18}.c) that helps to resolve topological
ambiguities, whenever a juncture actually is a POA.
The latter will be discussed further down below.

\subsection{Generation of a Single Tetrahedron}

\noindent
In this subsection, we shall demonstrate the extraction of an
iso-hypersurface for a single toxel (\textit{cf.}, Fig.~\ref{fig:04}.c).
We also indicate the relevant processing stages
(\textit{cf.}, Fig.~\ref{fig:03}).
\\
\indent
In $2$D, a single pixel is represented within a $2\times2$-neighborhood 
by a quarter of its (quadratic) area.
In $3$D, a single voxel is represented within a 
$2\times2\times2$-neighborhood by an eights of its (cubic) volume.

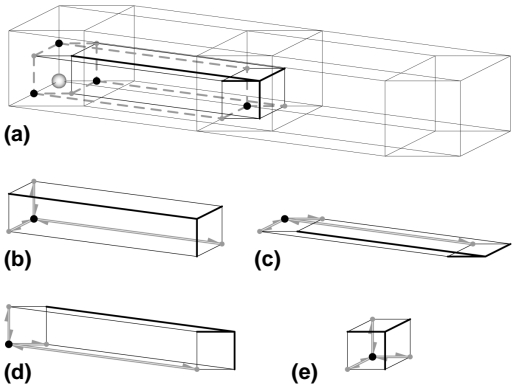
\begin{figure}[hb]
  \begin{center}
    \scalebox{0.58}{
    \epsfig{width=12.0cm,figure=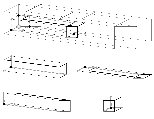} 
    \setlength{\unitlength}{1cm}
    \put(-12.19,5.81){\changefont{phv}{b}{n}\Large (a)}    
    \put(-12.19,2.94){\changefont{phv}{b}{n}\Large (b)}    
    \put(-6.45,2.94){\changefont{phv}{b}{n}\Large (c)}    
    \put(-12.19,0.35){\changefont{phv}{b}{n}\Large (d)}    
    \put(-5.60,0.35){\changefont{phv}{b}{n}\Large (e)}  
    }  
  \end{center}
  \vspace{-0.3cm} 
  \caption{
(a) A sixteenth of a toxel within a $4$D-cell at site no. $7$; 
detailed views of the corresponding boundary volume octants 
(b) no. $11 \:\oplus$;
(c) no. $9 \:\ominus$;
(d) no. $6 \:\ominus$;
(e) no. $18 \:\oplus$ (\textit{cf.}, Table~\ref{tab:02}).
} 
  \label{fig:08}
\end{figure}

\noindent
Analogously in $4$D, a single toxel is represented within a
$2\times2\times2\times2$-neighborhood by a sixteenth of its 
($4$-cubic) space-time \footnote{I.e., if we choose three spatial 
dimensions and a temporal one.}.
Hence, we need sixteen different $4$D-cells for the proper construction
of the complete hypersurface of a single toxel.
In order to demonstrate how STEVE uses the previously introduced
indexing scheme (\textit{cf.}, Fig.~\ref{fig:07}) in combination with 
the vector path table (\textit{cf.}, Table~\ref{tab:02}),
we shall process here a single sixteenth of a toxel.

\begin{figure}[t]
  \begin{center}
    \scalebox{0.58}{
    \epsfig{width=12.0cm,figure=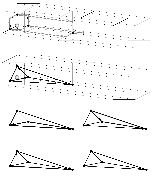} 
    \setlength{\unitlength}{1cm}
    \put(-12.08,10.70){\changefont{phv}{b}{n}\Large (a)}  
    \put(-12.08,6.68){\changefont{phv}{b}{n}\Large (b)}  
    \put(-12.08,3.52){\changefont{phv}{b}{n}\Large (c)}  
    \put(-6.34,3.52){\changefont{phv}{b}{n}\Large (d)}  
    \put(-12.08,0.35){\changefont{phv}{b}{n}\Large (e)}  
    \put(-6.34,0.35){\changefont{phv}{b}{n}\Large (f)}  
    }
  \end{center}
  \vspace{-0.3cm} 
  \caption{
Continuation of Fig.~\ref{fig:08}:
(a) the four initial cyclic vector paths within a $4$D-cell. 
The final tetrahedron together with the final cyclic vector path
(b) $11 \rightarrow 9 \rightarrow 6 \rightarrow 11$ (embedded into the
$4$D-cell); 
(c) $11 \rightarrow 6 \rightarrow 18 \rightarrow 11$;
(d) $18 \rightarrow 9 \rightarrow 11 \rightarrow 18$;
(e) $6 \rightarrow 9 \rightarrow 18 \rightarrow 6$.
(f) The final tetrahedron, where each of its edges represents
a pair of anti-parallel vectors.
} 
  \label{fig:09}
\end{figure}

\begin{figure}[b]
  \begin{center}
    \scalebox{0.58}{
    \epsfig{width=9.5cm,figure=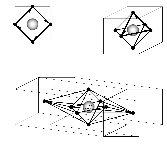} 
    \setlength{\unitlength}{1cm}
    \put(-9.82,5.08){\changefont{phv}{b}{n}\Large (a)}    
    \put(-5.45,5.08){\changefont{phv}{b}{n}\Large (b)}    
    \put(-9.82,0.36){\changefont{phv}{b}{n}\Large (c)}
    }    
  \end{center}
  \vspace{-0.3cm} 
  \caption{
Resulting manifolds of codimension $1$ for a single
(a) pixel in $2$D,
(b) voxel in $3$D, and 
(c) toxel in $4$D, respectively.
} 
  \label{fig:10}
\end{figure}

\indent
As an example, we shall consider within a $4$D-cell the four BVOs for 
the single toxel at site ID no.~$7$ (\textit{cf.},
Fig.~\ref{fig:08}.a).
These are the volumes 
no. $11 \:\oplus$ (in the positive $x$-direction; \textit{cf.}, 
Fig.~\ref{fig:08}.b),
no. $9 \:\ominus$ (in the negative $y$-direction; \textit{cf.}, 
Fig.~\ref{fig:08}.c),
no. $6 \:\ominus$ (in the negative $z$-direction; \textit{cf.}, 
Fig.~\ref{fig:08}.d),
and no. $18 \:\oplus$ (in the positive $t$-direction; \textit{cf.}, 
Fig.~\ref{fig:08}.e).
Hence, we have identified the initial paths (or TCV pairs; 
\textit{cf.}, Fig.~\ref{fig:03}.c).
In total we obtain $24$ initial vectors (\textit{cf.}, 
Fig.~\ref{fig:09}.a), 
which -- in a next step -- can be combined into the four initial 
cyclic (i.e., closed) vector paths, or the protomesh:
$37 \rightarrow 9 \rightarrow 34 \rightarrow 6 
\rightarrow 36 \rightarrow 11 \rightarrow 37$,  
$36 \rightarrow 6 \rightarrow 44 \rightarrow 18 
\rightarrow 49 \rightarrow 11 \rightarrow 36$, 
$47 \rightarrow 9 \rightarrow 37 \rightarrow 11 
\rightarrow 49 \rightarrow 18 \rightarrow 47$, and  
$34 \rightarrow 9 \rightarrow 47 \rightarrow 18 
\rightarrow 44 \rightarrow 6 \rightarrow 34$, respectively
(\textit{cf.}, Fig.~\ref{fig:03}.e).
\\
\indent
Note that at the junctures, TCVs are only connected such
that the end-point of a predecessor connects to the starting-point of 
a linked successor (which \textit{never} must be anti-parallel to the
preceding vector).
The vector connectivities at the boundary cube centers are always
predefined through the vector path table (\textit{cf.}, 
Table~\ref{tab:02}).
\\
\indent
In a further step, all junctures (i.e., those with point IDs 
above $31$, \textit{cf.}, Fig.~\ref{fig:07}.c) will be discarded
(\textit{cf.}, Fig.~\ref{fig:03}.f).
As a result, one obtains a single tetrahedron 
(\textit{cf.}, Fig.~\ref{fig:03}.h) that is represented
by its four final, reduced cyclic vector paths (\textit{cf.}, 
Fig.s~\ref{fig:09}.b--f).
Each of these cyclic vector paths represents a triangle, which is 
embedded into $4$D. 
Note that the initial orientations of the boundary volumes
have been passed on, such that a consistent evaluation of
$4$-normal vectors is possible (\textit{cf.}, Appendix~\ref{app:appA}).
\\
\indent
If we repeat the above processing for the remaining fifteen $4$D-cells, 
we end up in total with sixteen oriented tetrahedrons that in their 
union represent the final STEVE-hypersurface for a single, isolated 
active toxel (\textit{cf.}, Fig.~\ref{fig:10}.c).
The STEVE algorithm has stored at this stage the generated
$3$-simplices (\textit{cf.}, Fig.~\ref{fig:03}.j) and thereby
completes its processing.
Note that in Fig.~\ref{fig:10}.c the orientations of the triangles
of the tetrahedrons are not drawn, because in that 
$4$D space each triangle is oriented in both ways.
\\
\indent
For comparison we show in Fig.~\ref{fig:10}.a the DICONEX-contour 
-- consisting of four vectors -- for a single, isolated 
pixel~\cite{BRS09}. 
And in Fig.~\ref{fig:10}.b, we show the VESTA-surface -- consisting of 
eight oriented triangles -- for a single, isolated voxel~\cite{BRS12}. 
Note that each edge of the octahedron represents two anti-parallel 
vectors.

\begin{figure}[b]
  \begin{center}
    \scalebox{0.58}{
    \epsfig{width=12.0cm,figure=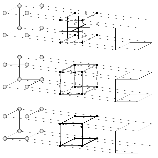} 
    \setlength{\unitlength}{1cm}
    \put(-11.91,8.52){\changefont{phv}{b}{n}\Large (a)}    
    \put(-11.91,4.60){\changefont{phv}{b}{n}\Large (b)}    
    \put(-11.91,0.68){\changefont{phv}{b}{n}\Large (c)} 
    }   
  \end{center}
  \vspace{-0.3cm} 
  \caption{
Generation of an isochronous hypersurface segment within a $4$D-cell:
(a) initial boundary volume octants;
(b) protomesh;
(c) final cube-shaped hypersurface segment, with one final 
cyclic (clockwise oriented) vector path indicating the orientation
of the hypersurface segment.
} 
  \label{fig:11}
\end{figure}

\subsection{Isochronous Hypersurface Segments}
\label{subsec:isochronous}

\noindent
While considering time as the fourth dimension, a so-called
isochronous hypersurface segment (\textit{cf.}, Ref.~\cite{HPET08},
for an application) may be generated. 
It has a rather simple geometry within a given $4$D-cell 
(i.e., similar to the generated single tetrahedrons in the previous 
subsection).
If all eight toxel sites of the ``past'' are active, and simultaneously
all eight toxel sites of the ``future'' are inactive (\textit{cf.}, 
Fig.~\ref{fig:11}), then one obtains as a result a cube-shaped segment 
(as shown in the figure at the intermediate time, here, the ``present''), 
which simply fills the whole $3$D space at the fixed time.
\\
\indent
Note that here the initial eight BVOs yield 
after consideration of both the indexing scheme (\textit{cf.},
Fig.~\ref{fig:07}) and the vector path table (\textit{cf.}, 
Table~\ref{tab:02}), six initial cyclic vector paths, i.e., 
the protomesh.
The orientations of the final vector paths are inherited from the
initial ones (i.e., either $\oplus$ or $\ominus$;
\textit{cf.}, Fig.s~\ref{fig:06}.e and~\ref{fig:06}.f, and 
Table~\ref{tab:02}) after the removal of the junctures.
In Fig.~\ref{fig:11}.b, the tiny black arrows mark the initial TCVs, 
which lead to the final vector $4$-cycles that 
are pronounced in Fig.~\ref{fig:11}.c.

\subsection{Subspaces and Bounding Shapes}

\noindent
In the previous two subsections, we have encountered two rather simple
$3$D shapes as resulting hypersurface segments, i.e., a tetrahedron and
a cube, respectively.
The $4$D-cells within which STEVE determines iso-hypersurface segments
are each bounded by eight $3$-cubes (or cubes), i.e., $3$D subspaces
(\textit{cf.}, Appendix~\ref{app:appB}). 
Both previously determined hypersurface segments, i.e., the tetrahedron
and the cube, are themselves bounded by either four triangles (i.e., 
$3$-cycles; \textit{cf.}, Fig.~\ref{fig:12}.3) or six squares (i.e., 
$4$-cycles; \textit{cf.}, Fig.~\ref{fig:12}.4.b), respectively.
Note that each triangle or square is embedded in a $3$D subspace.
\\
\indent
In fact, all possible hypersurface segments that the STEVE algorithm
will generate are bounded by the $N$-cycles
($N=3,4,5,6,7,8,9,12$), which VESTA~\cite{BRS12} would create, 
if it were processing the active voxels within properly arranged 
$3$D-cells (i.e., $2\times2\times2$-neighborhoods of voxels).
This observation agrees with the fact that the number of vector paths
(i.e., $192$) for the STEVE algorithm (\textit{cf.}, Table~\ref{tab:02}) 
equals to eight times of the number of vector paths (i.e.,
$8\times24$) for the marching variant of VESTA (\textit{cf.}, 
Ref.~\cite{BRS12}).
\\
\indent
In $2$D, the DICONEX~\cite{BRS09} algorithm determines all 
properly oriented line segments within the cube-bounding $2$D 
subspaces, i.e., its six squares.
In Fig.~\ref{fig:12}, the complete tiling sets of segments of the 
manifolds of codimension $1$ for $2$D and $3$D (sub)spaces as 
determined by the DICONEX, VESTA, and STEVE algorithms are pictured.
Note that multiple segments could be generated within in a given 
(sub)space (for more detail, \textit{cf.}, Ref.~\cite{BRS12}).

\subsection{Decomposition of Hypersurface Segments}

\noindent
This subsection addresses the decomposition of polytopes
within the STEVE algorithm (\textit{cf.}, Fig.~\ref{fig:03}.i).
E.g., for visualization purposes and/or for the purpose of
$4$-normal vector calculations (\textit{cf.}, Appendix~\ref{app:appA}) 
one has to decompose the polytopes into a set of 
tetrahedrons ($3$-simplices) \footnote{For more detail on the 
subject of polytope analysis, \textit{cf.}, Ref.~\cite{POLYMAKE}.}.

\begin{figure}[hb]
  \begin{center}
    \scalebox{0.58}{
    \epsfig{width=14.25cm,figure=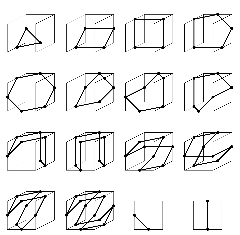} 
    \setlength{\unitlength}{1cm}
    \put(-14.53,13.29){\changefont{phv}{b}{n}\Large (3)}    
    \put(-11.02,13.29){\changefont{phv}{b}{n}\Large (4.a)}    
    \put(-7.50,13.29){\changefont{phv}{b}{n}\Large (4.b)}    
    \put(-3.99,13.29){\changefont{phv}{b}{n}\Large (5)}    
    \put(-14.53,9.78){\changefont{phv}{b}{n}\Large (6.a)}    
    \put(-11.02,9.78){\changefont{phv}{b}{n}\Large (6.b)}    
    \put(-7.50,9.78){\changefont{phv}{b}{n}\Large (6.c)}    
    \put(-3.99,9.78){\changefont{phv}{b}{n}\Large (6.d)}    
    \put(-14.53,6.26){\changefont{phv}{b}{n}\Large (7)}    
    \put(-11.02,6.26){\changefont{phv}{b}{n}\Large (8.a)}    
    \put(-7.50,6.26){\changefont{phv}{b}{n}\Large (8.b)}    
    \put(-3.99,6.26){\changefont{phv}{b}{n}\Large (8.c)}    
    \put(-14.53,2.75){\changefont{phv}{b}{n}\Large (9)}    
    \put(-11.02,2.75){\changefont{phv}{b}{n}\Large (12)}    
    \put(-7.50,2.75){\changefont{phv}{b}{n}\Large (2.a)}    
    \put(-3.99,2.75){\changefont{phv}{b}{n}\Large (2.b)} 
    }   
  \end{center}
  \vspace{-0.3cm} 
  \caption{
Bounding shapes (tiling sets) for DICONEX, VESTA, and STEVE:
(3) -- (12) fourteen different $3$D surface cycles for $3$D 
(sub)spaces;
(2.a) \& (2.b) two different $2$D contour segments for $2$D 
(sub)spaces, which can be combined into polygons of the previous 
$3$D-cells.
} 
  \label{fig:12}
\end{figure}

\begin{figure}[t]
  \begin{center}
    \scalebox{0.58}{
    \epsfig{width=11.5cm,figure=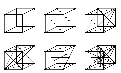} 
    \setlength{\unitlength}{1cm}
    \put(-11.77,6.29){\changefont{phv}{b}{n}\Large (a)}    
    \put(-7.905,6.29){\changefont{phv}{b}{n}\Large (b)}    
    \put(-4.04,6.29){\changefont{phv}{b}{n}\Large (c)}    
    \put(-11.77,2.77){\changefont{phv}{b}{n}\Large (d)}    
    \put(-7.905,2.77){\changefont{phv}{b}{n}\Large (e)}    
    \put(-4.04,2.77){\changefont{phv}{b}{n}\Large (f)} 
    }   
  \end{center}
  \vspace{-0.3cm} 
  \caption{
High resolution decomposition of a cube into twenty-four 
tetrahedrons (see text).
} 
  \label{fig:13}
\end{figure}

\begin{figure}[b]
  \begin{center}
    \scalebox{0.58}{
    \epsfig{width=14.25cm,figure=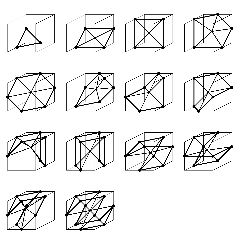} 
    \setlength{\unitlength}{1cm}
    \put(-14.53,13.29){\changefont{phv}{b}{n}\Large (3)}    
    \put(-11.02,13.29){\changefont{phv}{b}{n}\Large (4.a)}    
    \put(-7.50,13.29){\changefont{phv}{b}{n}\Large (4.b)}    
    \put(-3.99,13.29){\changefont{phv}{b}{n}\Large (5)}    
    \put(-14.53,9.78){\changefont{phv}{b}{n}\Large (6.a)}    
    \put(-11.02,9.78){\changefont{phv}{b}{n}\Large (6.b)}    
    \put(-7.50,9.78){\changefont{phv}{b}{n}\Large (6.c)}    
    \put(-3.99,9.78){\changefont{phv}{b}{n}\Large (6.d)}    
    \put(-14.53,6.26){\changefont{phv}{b}{n}\Large (7)}    
    \put(-11.02,6.26){\changefont{phv}{b}{n}\Large (8.a)}    
    \put(-7.50,6.26){\changefont{phv}{b}{n}\Large (8.b)}    
    \put(-3.99,6.26){\changefont{phv}{b}{n}\Large (8.c)}    
    \put(-14.53,2.75){\changefont{phv}{b}{n}\Large (9)}    
    \put(-11.02,2.75){\changefont{phv}{b}{n}\Large (12)} 
    }   
  \end{center}
  \vspace{-0.3cm} 
  \caption{
High resolution decomposition of the $3$D bounding shapes (tiling 
sets) for VESTA, and STEVE (\textit{cf.}, Fig.~\ref{fig:12}).
} 
  \label{fig:14}
\end{figure}

\begin{figure}[t]
  \begin{center}
    \scalebox{0.58}{
    \epsfig{width=15.25cm,figure=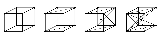} 
    \setlength{\unitlength}{1cm}
    \put(-15.52,2.76){\changefont{phv}{b}{n}\Large (a)}    
    \put(-11.66,2.76){\changefont{phv}{b}{n}\Large (b)}    
    \put(-7.80,2.76){\changefont{phv}{b}{n}\Large (c)}    
    \put(-3.94,2.76){\changefont{phv}{b}{n}\Large (d)}  
    }  
  \end{center}
  \vspace{-0.3cm} 
  \caption{
Low resolution decomposition of a cube into twelve 
tetrahedrons (see text).
} 
  \label{fig:15}
\end{figure}

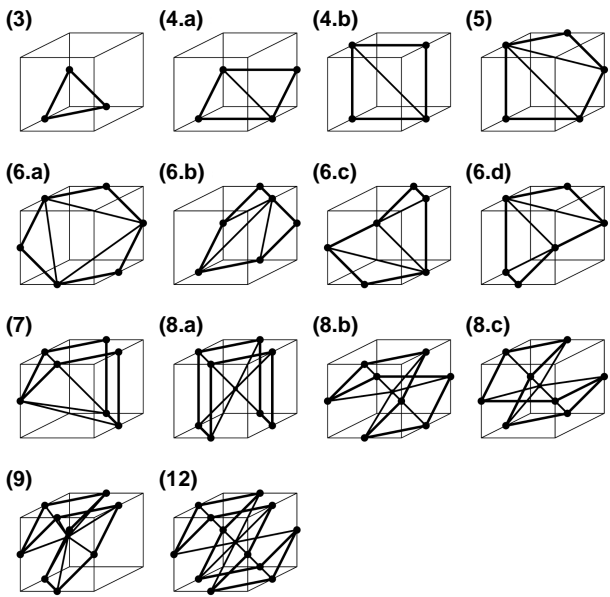
\begin{figure}[b]
  \begin{center}
    \scalebox{0.58}{
    \epsfig{width=14.25cm,figure=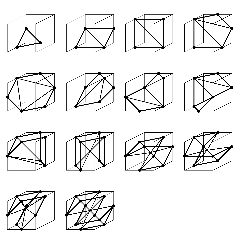} 
    \setlength{\unitlength}{1cm}
    \put(-14.53,13.29){\changefont{phv}{b}{n}\Large (3)}    
    \put(-11.02,13.29){\changefont{phv}{b}{n}\Large (4.a)}    
    \put(-7.50,13.29){\changefont{phv}{b}{n}\Large (4.b)}    
    \put(-3.99,13.29){\changefont{phv}{b}{n}\Large (5)}    
    \put(-14.53,9.78){\changefont{phv}{b}{n}\Large (6.a)}    
    \put(-11.02,9.78){\changefont{phv}{b}{n}\Large (6.b)}    
    \put(-7.50,9.78){\changefont{phv}{b}{n}\Large (6.c)}    
    \put(-3.99,9.78){\changefont{phv}{b}{n}\Large (6.d)}    
    \put(-14.53,6.26){\changefont{phv}{b}{n}\Large (7)}    
    \put(-11.02,6.26){\changefont{phv}{b}{n}\Large (8.a)}    
    \put(-7.50,6.26){\changefont{phv}{b}{n}\Large (8.b)}    
    \put(-3.99,6.26){\changefont{phv}{b}{n}\Large (8.c)}    
    \put(-14.53,2.75){\changefont{phv}{b}{n}\Large (9)}    
    \put(-11.02,2.75){\changefont{phv}{b}{n}\Large (12)} 
    }   
  \end{center}
  \vspace{-0.3cm} 
  \caption{
Low resolution decomposition of the $3$D bounding shapes (tiling 
sets) for VESTA, and STEVE (\textit{cf.}, Fig.~\ref{fig:12}).
} 
  \label{fig:16}
\end{figure}

\noindent
Due to the various $3$D bounding shapes as shown in 
Fig.s~\ref{fig:12}.3 --~\ref{fig:12}.12,
the polytopes generated by STEVE could be very complex-shaped.
\\
\indent
As an example, we demonstrate in Fig.~\ref{fig:13}, how to 
decompose a single cube into twenty-four tetrahedrons.
Fig.~\ref{fig:13}.a shows a cube, which consists of six $3$D tilings
($4$-cycles) as shown in Fig.~\ref{fig:12}.4.b.
In Fig.~\ref{fig:13}.b, for each $4$-cycle its center of mass point 
(face center) is marked.
In Fig.~\ref{fig:13}.c, the surface cycles are decomposed into 
triangles while connecting the face centers with the corresponding 
$4$-cycle support points.
Each newly drawn line actually represents a pair of anti-parallel 
vectors.
Note that neither triangles nor single tetrahedrons 
will be decomposed by us any further~\footnote{However, this is 
done so, e.g., in Ref.~\cite{HUOV12}.}.
\\
\indent
Next, in Fig.~\ref{fig:13}.d, all triangles that enclose a particular 
single volume are collected into an object along with the information, 
to which $N$-cycles the triangles belong; 
furthermore, the absolute center of the enclosed volume is determined.
In Fig.~\ref{fig:13}.e, lines are introduced that connect the face 
centers with the absolute volume center.
Finally, in Fig.~\ref{fig:13}.f, the additional connections of all 
$4$-cycle support points with the absolute volume center finally yields
the twenty-four properly oriented tetrahedrons.
In Fig.~\ref{fig:14}, we show the high resolution decompositions
of all possible $3$D tiling sets.
\\
\indent
As an alternative, we show in Fig.~\ref{fig:15} a lower resolution
decomposition that is similar to the previous one.
In particular, the steps that are shown in Fig.s~\ref{fig:13}.b 
and ~\ref{fig:13}.c are replaced by $N$-cycle decompositions,
which do not have any additional center of mass points
(\textit{cf.}, Ref.~\ref{fig:15}.b).
Hence, in Fig.~\ref{fig:15} the step that is shown in 
Fig.~\ref{fig:13}.e is omitted.
This latter procedure only yields twelve, but also properly oriented 
tetrahedrons.
In Fig.~\ref{fig:16}, we show the low resolution decompositions
of all possible $3$D tiling sets.
\\
\indent
Within STEVE, one can use both -- i.e., high resolution (HR),
and low resolution (LR) -- types of decompositions into $3$-simplices 
for the potentially complex shaped polytopes.
Note that the polytopes, which are embedded into $4$D,
could be decomposed also while using fewer tetrahedrons.
E.g., one could decompose a cube into just five tetrahedrons.
However, the usage of fewer tetrahedrons could introduce stronger
directional dependencies into the hypersurface (\textit{cf.}, 
application subsection~\ref{subsec:highlow}).

\begin{figure}[t]
  \begin{center}
    \scalebox{0.58}{
    \epsfig{width=12.0cm,figure=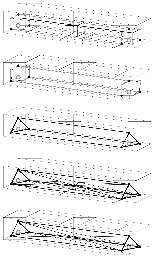} 
    \setlength{\unitlength}{1cm}
    \put(-12.01,16.48){\changefont{phv}{b}{n}\Large (a)}    
    \put(-12.01,12.51){\changefont{phv}{b}{n}\Large (b)}    
    \put(-12.01,8.54){\changefont{phv}{b}{n}\Large (c)}    
    \put(-12.01,4.58){\changefont{phv}{b}{n}\Large (d)}    
    \put(-12.01,0.61){\changefont{phv}{b}{n}\Large (e)}
    }    
  \end{center}
  \vspace{-0.3cm} 
  \caption{
Generation of a triangular strut within a $4$D-cell (see text).
} 
  \label{fig:17}
\end{figure}
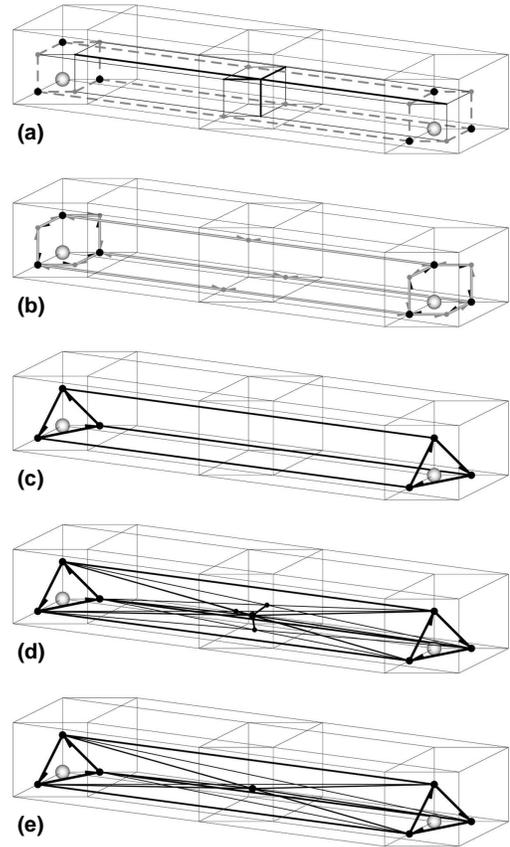

\subsection{Generation of a Triangular Strut}

\noindent
In another example, we briefly demonstrate the generation of a
hypersurface segment that has the shape of a triangular strut.
In Fig.~\ref{fig:17}, two toxels (in NV) at site IDs no.~$7$ 
and no.~$15$ are in contact through a single volume.
This volume of contact is visible within the cube that represents 
here the ``present'' (\textit{cf.}, Fig.~\ref{fig:17}.a).
It actually is made up by the superposition of two BVOs that each have 
opposite orientations, i.e., no.s $18 \:\oplus$ and $18 \:\ominus$ 
(in the positive and negative $t$-directions, respectively). 
Hence, their TCVs \textit{cancel out} each one another, and there is no
contribution to the protomesh from this volume.
\\
\indent
Once again, the application of both, the indexing scheme 
(\textit{cf.}, Fig.~\ref{fig:07}) and the vector path table 
(\textit{cf.}, Table~\ref{tab:02}), yields five initial cyclic 
vector paths (\textit{cf.}, Fig.~\ref{fig:17}.b), i.e., the protomesh.
Fig.~\ref{fig:17}.c shows the final trianglular strut-shaped 
hypersurface segment, with two final cyclic vector paths indicating 
its orientation.
Note that this segment is bounded by two $3$-cycles as 
shown in Fig.~\ref{fig:12}.3, and by three $4$-cycles as shown in 
Fig.~\ref{fig:12}.3.a, respectively.
Finally, in Fig.s~\ref{fig:17}.d and~\ref{fig:17}.e, 
we picture HR- and LR-decompositions of the hypersurface segment 
into fourteen and eight oriented tetrahedrons, respectively 
(\textit{cf.}, Fig.s~\ref{fig:13} and~\ref{fig:15}).
\\
\noindent
For this particular example, the STEVE algorithm made use of all 
processing steps that are indicated in the flowchart (\textit{cf.}, 
Fig.~\ref{fig:03}) except for the support point interpolation and
the identification of POAs (\textit{cf.}, Fig.s~\ref{fig:03}.g
and~\ref{fig:03}.d, respectively).
Since we did not yet encounter any particular topological ambiguities
while constructing hypersurface segments, we are going to discuss 
their treatment next.

\begin{figure}[t]
  \begin{center}
    \scalebox{0.58}{
    \epsfig{width=12.0cm,figure=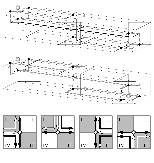} 
    \setlength{\unitlength}{1cm}
    \put(-12.04,8.97){\changefont{phv}{b}{n}\Large (a)}    
    \put(-12.04,4.96){\changefont{phv}{b}{n}\Large (b)}    
    \put(-12.04,0.36){\changefont{phv}{b}{n}\Large (c)}
    }    
  \end{center}
  \vspace{-0.3cm} 
  \caption{
Encounter of a topological ambiguity within a $4$D-cell:
(a) initial boundary cubes;
(b) protomesh;
(c) connectivity diagram.
} 
  \label{fig:18}
\end{figure}

\subsection{Ambiguous Connectivity}

\noindent
The discretized spaces that we consider here could lead to 
topological ambiguities.
In $2$D this is the case, if two active pixels share only one
common point (i.e., a vertex; \textit{cf.}, Ref.~\cite{BRS09} 
for more detail).
In $3$D this is the case, if two active voxels share only one
common edge (\textit{cf.}, Ref.~\cite{BRS12} for more detail).
Analogously, this is the case in $4$D, if two active toxels
share only one common face (square).
\\
\indent
In Fig.~\ref{fig:18}.a, we show a $4$D-cell with two active toxels
at site IDs No.~$4$ and No.~$15$.
The (sixteenth of the) toxels are in contact through a single face.
Within the cube that here represents the ``present'' this surface
of contact is visible.
It actually is made up by the contact of the two neighboring BVOs
no.s $16 \:\oplus$ and $18 \:\ominus$
(in the positive and negative $t$-directions, respectively). 
In particular for both BVOs, their paths no.s $1$ and $3$ (\textit{cf.},
Table~\ref{tab:02}) start and end at juncture no. $47$, 
respectively (\textit{cf.}, Fig~\ref{fig:03}.d).
\\
\indent
Hence, we have in Fig.~\ref{fig:18}.b an ambiguous configuration
(white dot) for the protomesh.
Evidently, this particular juncture has turned into a so-called 
``point of ambiguity'' (POA; \textit{cf.}, Ref.~\cite{BRS12}).
The connectivity diagram (\textit{cf.}, Fig.~\ref{fig:18}.c)
helps to consistently resolve which preceding TCV should connect
to a succeeding TCV.
Each incoming (white, straight) vector can connect to either
one of the two (black, straight) vectors, depending on the chosen
connectivity mode.
Using the field values for the toxels at positions \textit{I},
\textit{II}, \textit{III}, and \textit{IV} in the figure, 
one may assign their average field value to the POA.
\\
\indent
E.g., for an average field value below (above) the desired 
iso-value of the hypersurface, one generates the local ``disconnect'' 
(``connect'') mode while pursuing the white (black) bent directed 
path for each incoming vector consistently.
This latter treatment (also known as ``mixed'' mode~\cite{BRS12} 
treatment) will allow for an automated, robust resolution of all 
encountered ambiguities.
Note that one may enforce the connectivity modes also globally
onto the complete considered $4$D data set, by either \textit{always} 
selecting the ``disconnect'', or the ``connect'' mode.
\\
\noindent
In Fig.~\ref{fig:19}.a, we show two final hypersurface segments 
(tetrahedrons) that result from the ``disconnect'' mode.
Alternatively, we show in Fig.~\ref{fig:19}.b one final hypersurface 
segment that results from the ``connect'' mode.
This single segment is bounded by four $3$-cycles as 
shown in Fig.~\ref{fig:12}.3, and by two $6$-cycles as shown in 
Fig.~\ref{fig:12}.6.b, respectively.
In Fig.~\ref{fig:19}.c, it has been HR-decomposed (\textit{cf.},
Fig.~\ref{fig:13}) into sixteen oriented tetrahedrons, and in
Fig.~\ref{fig:19}.d, it has been LR-decomposed (\textit{cf.},
Fig.~\ref{fig:15}) into just twelve oriented tetrahedrons.
Note that each drawing in Fig.~\ref{fig:19} shows two final cyclic 
vector paths indicating the orientation of the hypersurface segments.
\\
\indent
We would like to stress that due to the here considered
\textit{discretized} $4$D spaces, STEVE can generate \textit{six 
different} types of hypersurfaces: 
both high (i.e., HR) and low (i.e., LR) resolution hypersurfaces,
each one of them in either ``mixed'', or global ``disconnect'' and
``connect'' modes. 

\begin{figure}[t]
  \begin{center}
    \scalebox{0.58}{
    \epsfig{width=12.0cm,figure=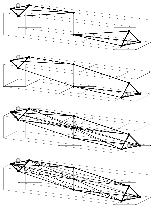} 
    \setlength{\unitlength}{1cm}
    \put(-12.01,12.49){\changefont{phv}{b}{n}\Large (a)}    
    \put(-12.01,8.52){\changefont{phv}{b}{n}\Large (b)}    
    \put(-12.01,4.55){\changefont{phv}{b}{n}\Large (c)}    
    \put(-12.01,0.59){\changefont{phv}{b}{n}\Large (d)}
    }    
  \end{center}
  \vspace{-0.3cm} 
  \caption{
Continuation of Fig.~\ref{fig:18}:
(a) two resulting tetrahedrons from the ``disconnect'' mode;
(b) a final, more complex-shaped hypersurface segment, 
resulting from the ``connect'' mode;
(c) as in (b), but with a hypersurface segment decomposisiton into 
sixteen tetrahedrons.
(d) as in (c), but with a decomposition into twelve tetrahedrons only.
} 
  \label{fig:19}
\end{figure}

\begin{figure}[t]
  \begin{center}
    \scalebox{0.58}{
    \epsfig{width=12.0cm,figure=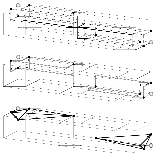} 
    \setlength{\unitlength}{1cm}
    \put(-12.00,8.77){\changefont{phv}{b}{n}\Large (a)}    
    \put(-12.00,4.75){\changefont{phv}{b}{n}\Large (b)}    
    \put(-12.00,0.73){\changefont{phv}{b}{n}\Large (c)} 
    }   
  \end{center}
  \vspace{-0.3cm} 
  \caption{
Hypersurface segment generation for two toxels, which are in
contact through a single edge:
(a) initial boundary cubes;
(b) two disjunct protomeshes;
(c) two resulting tetrahedrons.
} 
  \label{fig:20}
\end{figure}

\begin{figure}[t]
  \begin{center}
    \scalebox{0.58}{
    \epsfig{width=12.0cm,figure=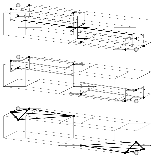} 
    \setlength{\unitlength}{1cm}
    \put(-12.00,8.77){\changefont{phv}{b}{n}\Large (a)}    
    \put(-12.00,4.75){\changefont{phv}{b}{n}\Large (b)}    
    \put(-12.00,0.73){\changefont{phv}{b}{n}\Large (c)}
    }    
  \end{center}
  \vspace{-0.3cm} 
  \caption{
Hypersurface segment generation for two toxels, which are in
contact through a single vertex:
(a) initial boundary cubes;
(b) two disjunct protomeshes;
(c) two resulting tetrahedrons.
} 
  \label{fig:21}
\end{figure}

\subsection{Disjunct Hypersurface Segments}

\noindent
In $3$D, VESTA never connects two voxels that are in contact
only through a single vertex (\textit{cf.}, Ref.~\cite{BRS12}
for more detail).
In $4$D, we have a similar situation when considering toxels that are 
in direct contact.
\\
\indent
In Fig.s~\ref{fig:20} and~\ref{fig:21}, the process of hypersurface 
segment generation (\textit{cf.}, Ref.~\ref{fig:03}) is shown for 
two toxels that are in contact only through a single edge and through 
a single vertex, respectively.
In both cases, we simply obtain two tetrahedrons as final hypersurface
segments, since the initial cyclic vector paths form two disjunct sets 
with four cyclic paths each.
Apparently, toxel pairs with such a weak connectivity will always
result in two separate hypersurface segments.
This concludes the technical section of this paper.

\section{Applications}
\label{sec:applications}

\noindent
In this application section, we present various examples for 
hypersurface generation in $4$D.
\\
\indent
These are (i) the shape characterization of a $4$D transversal phase 
space in the field of ion beam transport, (ii) the continuous 
transmutation of a square into two triangles, (iii) the isotherme 
evolution of a fireball expansion in the field of relativistic 
heavy-ion physics, and (iv) the study of intersections of low and high 
resolution hypersurfaces for a shrinking sphere.

\begin{figure}[t]
  \begin{center}
    \scalebox{0.58}{
    \epsfig{width=14.0cm,figure=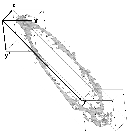} 
    }
  \end{center}
  \vspace{-0.3cm} 
  \caption{
Projection of the full $4$D hypersurfaces (i.e., volumes) generated 
by STEVE in the $4$D transversal phase space of a charged 
particle beam (see text).
} 
  \label{fig:22}
\end{figure}

\subsection{$4$D Shape Characterization}

\noindent
In the field of accelerator physics, for charged particle beams both 
tomographic reconstructions (\textit{cf.}, e.g., Ref.~\cite{HOCK13}) and
simulation data (\textit{cf.}, e.g., Ref.~\cite{SPAE14}) of the $4$D 
transverse phase space are available.
Accelerator physicists characterize the beams typically by using beam 
ellipsoids (\textit{cf.}, e.g., Ref.~\cite{CHAO99}), while assuming
that there is a decoupling between the radial and axial components in 
the transverse plane.
However, in reality this assumption may be unwarranted because the 
occupied phase space may have a quite different shape due to 
non-linear effects.
STEVE can help to visualize the particular phase space shape.
\\
\indent
In Fig.~\ref{fig:22}, we show a projection of the full hypersurfaces 
that result from the application of the STEVE algorithm to the 
$4$D phase space simulation data that have been presented 
in Ref.~\cite{SPAE14}.
STEVE generates for this $50\times50\times50\times50$-sized toxel data 
set with an iso-value of $1$ (i.e., it is enclosing all entries larger 
than $0$ in the grid) $129,563$ tetrahedrons in its high resolution 
``mixed'' mode.
In the figure, $x$ and $y$ denote the radial and axial deviations 
from the ideal beam position in the transversal plane, whereas
$x^\prime$ and $y^\prime$ denote the corresponding angles, 
respectively.
Interpolations have \textit{not} been applied here to the 
$102,037$ $4$D support points.

\subsection{Bifurcation within a $4$-Cell}

\begin{figure}[hb]
  \begin{center}
    \scalebox{0.58}{
    \epsfig{width=11.0cm,figure=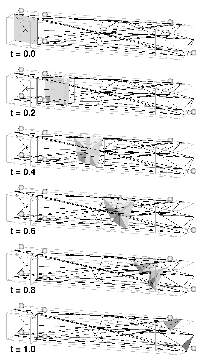} 
    }
  \end{center}
  \vspace{-0.3cm} 
  \caption{
Continuous transformation of a single quadrilateral into two 
separate triangles (see text).
} 
  \label{fig:23}
\end{figure}

\noindent
Typically, the meshes of tetrahedrons that represent the
final hypersurfaces are very dense.
In order to learn more about their inherent shape features, it is 
useful to intersect the generated manifolds with a $3$D space at 
a certain fixed value.
E.g., one could keep the fourth dimension (which we have named time
here) at a particular fixed value and then visualize the resulting
shape.
Note, that such shapes could consist of single points, line segments,
triangles, quadrilaterals, and/or (the whole volume of intersected) 
tetrahedrons (\textit{cf.}, the isochronous hypersurface segments, 
above).
\\
\indent
In Fig.~\ref{fig:23}, we show a sequence of one and the same $4$D-cell 
(i.e., in NV) with six activated toxel sites each.
In the ``past'' the four toxels with IDs $4,5,6$ and $7$ 
are active, whereas in the ``future'' the two toxels 
with IDs $12$ and $14$ are active (\textit{cf.}, Fig.~\ref{fig:07}.a).
STEVE has been applied to this configuration in its global high
resolution ``disconnect'' mode, in order to determine the hypersurface 
section (i.e., the network of black lines in the figure).
For each tesseract (except for the first and the last one), an 
additional cube is drawn for various fixed times $t$.
The parameter, $t$, denotes a relative value that ranges
between the extremes of $0$ and $1$ for intersections.\\
\indent
For the generated hypersurface segment, Fig.~\ref{fig:23} shows an
evolution of surface segments as a result of the chosen intersections.
Here it is demonstrated, how a single square may transform
continuously and smoothly into two separated triangles.
The pictured $4$D hypersurface segment (i.e., $32$ tetrahedrons) 
establishes a correspondence between the surfaces in the ``past'' 
and in the ``future'' (and between those anywhere in between).

\subsection{Isotherme Evolution}
\label{subsec:evolution}

\noindent
The quest for the correct equation of state of nuclear matter
continues (\textit{cf.}, e.g., Ref.~\cite{FRIM10} and Ref.s therein).
E.g., relativistic hydrodynamical models that simulate
relativistic heavy ion collisions are employed for its particular
characterization~\cite{STRO86,CSER94}.
In doing so, one may be faced with the task of freezeout hypersurface 
(FOHS) extraction from $3+1$D hydrodynamic simulation data 
(\textit{cf.}, e.g., Ref.~\cite{CHEN10} and Ref.s therein). 
In this example, a relativistic fluid has been propagated numerically 
on a cartesian $3$D grid.
\\

\begin{figure}[t]
  \begin{center}
    \scalebox{0.58}{
    \epsfig{width=15.0cm,figure=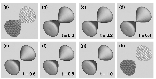} 
    }
  \end{center}
  \vspace{-0.3cm} 
  \caption{
Generation of chronologically developing isothermes:
(a) \& (h) selected grid cells of the discretized $3$D temperature 
fields at two subsequent time steps;
(b) -- (g) various temporal hypersurface intersections.
} 
  \label{fig:24}
\end{figure}

\noindent
The spheres in Fig.s~\ref{fig:24}.a and~\ref{fig:24}.h represent 
grid cells above a certain threshold temperature at two 
subsequent time steps (in GV) for the here considered fireball of hot,
dense nuclear matter that decays into two pieces.
For visualization purposes, STEVE has been used (in its high 
resolution global ``disconnect'' mode) in order to determine the 
(continuous) isothermal hypersurface that lies between the two 
shown $3$D data sets.
Various temporal intersections within the bounds given by 
Fig.s~\ref{fig:24}.a and~\ref{fig:24}.h are shown here for 
illustration.
\\
\indent
Note that the $4$D support point set has been interpolated linearly.
In particular, STEVE provides interpolations for all other
field values that have been associated with each toxel as well.
For the purpose of FOHS extraction, STEVE will not just be applied 
to two subsequent time steps as shown in the figure, but to any full 
set of $4$D simulation data.
The resulting total FOHS will allow for calculations of so-called 
observables (\textit{cf.}, e.g., Ref.~\cite{BRS97}).

\subsection{High vs. Low Resolution Rendering}
\label{subsec:highlow}

\begin{figure}[t]
  \begin{center}
    \scalebox{0.58}{
    \epsfig{width=15.0cm,figure=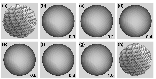}
    } 
  \end{center}
  \vspace{-0.3cm} 
  \caption{
A shrinking sphere:
(a) \& (h) selected grid cells for two, i.e., initial and final, 
$3$D data sets; 
(b) -- (g) various intersections of the generated 
high resolution iso-hypersurface;
the numbers refer to the intersection parameter, $t$.
} 
  \label{fig:25}
\end{figure}

\begin{figure}[t]
  \begin{center}
    \scalebox{0.58}{
    \epsfig{width=15.0cm,figure=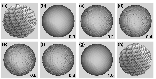} 
    }
  \end{center}
  \vspace{-0.3cm} 
  \caption{
As in Fig.~\ref{fig:25}, but for a generated iso-hypersurface
of lower resolution.
Note the additional artifacts at intersections (c) -- (f).
} 
  \label{fig:26}
\end{figure}

\noindent
In this subsection, we discuss the effects of choosing a lower 
resolution for the decomposition (\textit{cf.}, Fig.s~\ref{fig:15}
and~\ref{fig:16} vs. Fig.s~\ref{fig:13} and~\ref{fig:14}) 
of the polytopes generated by STEVE.
In Fig.s~\ref{fig:25} and~\ref{fig:26}, we each show the same two 
pairs of (i.e., without any noise) constructed volumetric data.
Both sets of data represent massive spheres, where the second
sphere has a slightly smaller radius.
For the union of the two $3$D data sets (i.e., $4$D data) 
both a HR- and a LR-hypersurface have been generated with STEVE.
\\
\indent
These have been intersected in the same manner as the FOHS
segment of the previous subsection.
Note that the lower resolution decomposition leads to additional
artifacts in the resulting surfaces (\textit{cf.}, 
Fig.s~\ref{fig:26}.c through~\ref{fig:26}.f).
We would like to emphasize that, e.g., a $4$D generalization of the
(extended) Marching Cubes algorithm~\cite{LORE87,BOUR94} would
generate such artifacts, because it provides only lower resolution 
surface templates.

\begin{table}[b]
\caption{
Number of directed paths vs. number of templates as a function of
integral dimension, $N$
}
\label{tab:03}
\begin{center}
\begin{tabular}{rlrlr}
\toprule
&\multicolumn{2}{c}{\textbf{No. of Paths}}
&\multicolumn{2}{c}{\textbf{No. of Templates}}\\
\addlinespace[0.5ex]
\toprule
{\boldmath$3$}\textbf{D}$\:$
&\textbf{(VESTA)} &$\: 24 \:$
&\quad\textbf{(MCA)} &$\: 256 \:$\\
{\boldmath$4$}\textbf{D}$\:$
&\textbf{(STEVE)} &$\: 192 \:$
&\multicolumn{2}{r}{$\: 65,536\:$}\\
{\boldmath$5$}\textbf{D}$\:$
&\multicolumn{2}{r}{$\: 1920 \:$}
&\multicolumn{2}{r}{$\: 4,294,967,296 \:$}\\
{\boldmath$6$}\textbf{D}$\:$
&\multicolumn{2}{r}{$\: 23,040 \:$}
&\multicolumn{2}{r}{$\: \approx\:1.845\times 10^{19} \:$}\\
{\boldmath$7$}\textbf{D}$\:$
&\multicolumn{2}{r}{$\: 322,560 \:$}
&\multicolumn{2}{r}{$\: \approx\:3.403\times 10^{38} \:$}\\
{\boldmath$8$}\textbf{D}$\:$
&\multicolumn{2}{r}{$\: 5,160,960 \:$}
&\multicolumn{2}{r}{$\: \approx\:1.158\times 10^{77} \:$}\\
{\boldmath$9$}\textbf{D}$\:$
&\multicolumn{2}{r}{$\: 92,897,280 \:$}
&\multicolumn{2}{r}{$\: \approx\:1.341\times 10^{154} \:$}\\
{\boldmath$10$}\textbf{D}$\:$
&\multicolumn{2}{r}{$\: 1,857,945,600 \:$}
&\multicolumn{2}{r}{$\: \approx\:1.798\times 10^{308} \:$}\\
{\boldmath$11$}\textbf{D}$\:$
&\multicolumn{2}{r}{$\: 40,874,803,200 \:$}
&\multicolumn{2}{r}{$\: \approx\:3.232\times 10^{616} \:$}\\
\addlinespace[0.8ex]
{\boldmath$N$}\textbf{D}$\:$
&\multicolumn{2}{c}{$\: \displaystyle \prod_{k\:=\:2}^N 2k \:$}
&\multicolumn{2}{c}{\raisebox{-0.4ex}
{\scalebox{1.4}{$\: \displaystyle 2^{2^N} \:$}}}\\
\addlinespace[0.5ex]
\bottomrule
\end{tabular}
\end{center}
\textbf{Note:} The estimated number of atoms in our universe is 
$\sim 10^{80}$ (\textit{cf.}, Ref.~\cite{WOLFRAM}).
\end{table}

\section{Implementation Issues}
\noindent
The original Marching Cubes algorithm (MCA) by Lorenzen and 
Cline~\cite{LORE87} used only fourteen templates for the construction
of surfaces from $3$D volumetric images.
This algorithm had to be extended, because the number of
templates was \textit{insufficient}, i.e., it could not warrant 
final surfaces that do not contain any holes.
An implementation of the extended MCA can be found in 
Ref.~\cite{BOUR94}.
However, in order to obtain an efficient look-up table based computer 
program, $256$ configurations (\textit{cf.}, Table~\ref{tab:03}) 
-- rather than only $17$ -- are implemented.
\\
\indent
In contrast, VESTA only requires the storage of half of the
$24$ directed paths (\textit{cf.}, Table~\ref{tab:01}), due to 
its $\oplus / \ominus$-symmetry.
Note that the extended MCA generates the exact surfaces as VESTA,
if the latter is executed in its low-resolution ``disconnect''
mode~\cite{BRS12}.
I.e., unlike VESTA the MCA \textit{cannot} be considered data-driven, 
because it cannot deal with a ``mixed'' connectivity mode.
E.g., the MCA cannot produce the $3$D surface tiles as shown in
Fig.s~\ref{fig:16}.8.a --~\ref{fig:16}.12.
\\
\indent
The authors of Ref.~\cite{BHAN04} state that they only require
$222$ configurations to provide templates for building hypersurfaces
in $4$D.
However, these $222$ configurations are used to fill a 
$2^{16}$-sized (i.e., $65,536$-sized) look-up table, 
whereas STEVE only requires
the storage of half of the $192$ directed paths (\textit{cf.}, 
Tables~\ref{tab:02} and~\ref{tab:03}), due to its 
$\oplus / \ominus$-symmetry.
\\
\indent
If we consider even higher dimensions, due to Table~\ref{tab:03} it would be 
quite impossible to provide a template based hypersurface generator 
-- in the strict sense -- for instance for a $11$-dimensional M-Theory, 
whereas our technology~\cite{PAT11} would not run into
resource problems that easily.

\section{Summary}

\noindent
In summary, the $4$D protomesh-based iso-hypersurface construction
algorithm STEVE has been described here for the very first time in 
great detail.
The STEVE algorithm basically requires the storage of $192$ vector paths only.
A minimum of data redundancy is achieved by avoiding the multiple counting of 
the manifold supporting points within the $4$D-cells.
\\
\indent
We would like to stress that there is \textit{more than one} iso-hypersurface 
solution.
This fact has not been pointed out explicitly by other authors yet.
It is a consequence of the discretization of the $4$D spaces 
under consideration, where topological ambiguities will allow for 
two different global treatments, 
and/or an additional local treatment, which is purely data-driven.
The proper treatment of possible ambiguities warrants that accidental 
rifts in the final hypersurfaces \textit{cannot be created}.
\\
\indent
STEVE constructs iso-hypersurfaces while using a \textit{single} 
$4$D building block (\textit{cf.}, e.g., Fig.~\ref{fig:06}.d).
It continuously transforms the set of initially identified
octants of boundary cubes into the final set of tetrahedrons.
One can choose between both a high and a lower resolution 
decomposition of the generated polytopes.
This final set of tetrahedrons represents one or more hypersurfaces.
Note that the initially given information about the interior/exterior
of the enclosed $4$D regions will be propagated to the final results.

\section*{Acknowledgements}

\noindent
Some initial work has been supported by the Department of
Energy under contract W-7405-ENG-36.
At the time, the particular funding has resulted in a first 
version of ANSI C-based software~\cite{STEVE04}, where ambiguities 
are resolved globally only.
The author wishes to thank the organizers of the mini workshop on
``Freeze-Out'', which was held at the Institute for Theoretical 
Physics, Goethe-University of Frankfurt, Germany, on May $6^{th}$, 
2009, for the opportunity to give a talk on this research.

\appendices

\section{Path Orientations}
\label{app:appA}

\begin{figure}[b]
  \begin{center}
    \scalebox{0.58}{
    \epsfig{width=9.5cm,figure=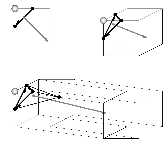} 
    \setlength{\unitlength}{1cm}
    \put(-9.82,5.08){\changefont{phv}{b}{n}\Large (a)}    
    \put(-5.45,5.08){\changefont{phv}{b}{n}\Large (b)}    
    \put(-9.82,0.36){\changefont{phv}{b}{n}\Large (c)}
    }    
  \end{center}
  \vspace{-0.3cm} 
  \caption{
Normalized normal vectors (gray) that are attached to the centers of 
the oriented ($N-1$)-simplices, which enclose the active $N$-dimensional
picture element sites with ID $0$ within the shown $N$D-cells:
(a) in $2$D (b) in $3$D, and (c) in $4$D, respectively.
} 
  \label{fig:27}
\end{figure}

\noindent
In Fig.~\ref{fig:27}, we show normalized normal vectors in $2$D,
$3$D, and $4$D, respectively, which are attached to the centers of
selected oriented ($N-1$)-simplices.
Each one of the simplices encloses an active $N$-dimensional picture
element site with ID $0$ within the shown 
$2(\times2)^{N-1}$-neighborhoods.
In particular, the normalized normal vectors point to the exterior 
of the enclosed $N$-dimensional regions, i.e., each component of
the normal vectors is larger than zero.
The orientations of normal vectors depend on the orientations
of the corresponding simplices, and vice versa.
Hence, the orientations of the initial protomesh building block vectors 
depend on the orientations of the normal vectors.
\\
\indent
Let the coordinate systems be chosen here as shown in Fig.~\ref{fig:04}.
The demand that all components of the normal vectors should
be larger than zero leads to the orientation of the contour vector
(black arrow) in $2$D and the orientation of the triangle (triplet of
black vectors) in $3$D as shown in Fig.s~\ref{fig:27}.a 
and~\ref{fig:27}.b, respectively.
In particular, the $\oplus$/$\ominus$ naming convention
of Table~\ref{tab:01} becomes apparent for $3$D.
In $4$D (similar to the lower dimensional cases), the particular choices 
depend on the way a $4$D normal (or $4$-normal) vector is constructed.

\indent
In application subsection~\ref{subsec:evolution}, $4$D space-time plays 
a vital role.
Therefore, we consider first the more complex situation of relativity.
Let $x^\mu = (x, y, z, t)$ be a contravariant event with the three 
spatial cartesian components $x$, $y$, and $z$, and the time $t$, 
respectively.
I.e.,
\begin{equation}
	x^\mu = x\:{\bf e}_{x} + y\:{\bf e}_{y} 
              + z\:{\bf e}_{z} + t\:{\bf e}_{t} \enspace ,
\end{equation}

\noindent
while considering
the \textit{canonical} multilinear basis
\begin{eqnarray}
&&\lbrace {\bf E}_{i} \rbrace = \lbrace\:1 \enspace, \enspace 
{\bf e}_{x} \enspace, \enspace {\bf e}_{y} \enspace, \enspace 
{\bf e}_{z} \enspace, \enspace {\bf e}_{t} \enspace, \enspace
{\bf e}_{x} \wedge {\bf e}_{y} \enspace, \nonumber\\ 
&&\quad{\bf e}_{x} \wedge {\bf e}_{z} \enspace, \enspace
{\bf e}_{x} \wedge {\bf e}_{t} \enspace, \enspace
{\bf e}_{y} \wedge {\bf e}_{z} \enspace, \enspace 
{\bf e}_{y} \wedge {\bf e}_{t} \enspace, \enspace 
{\bf e}_{z} \wedge {\bf e}_{t} \enspace, \nonumber\\
&&\quad{\bf e}_{x} \wedge {\bf e}_{y} \wedge {\bf e}_{z} 
\enspace, \enspace
{\bf e}_{x} \wedge {\bf e}_{y} \wedge {\bf e}_{t} \enspace, \enspace
{\bf e}_{x} \wedge {\bf e}_{z} \wedge {\bf e}_{t} \enspace, \nonumber\\
&&\quad{\bf e}_{y} \wedge {\bf e}_{z} \wedge {\bf e}_{t} \enspace, \enspace
{\bf e}_{x} \wedge {\bf e}_{y} \wedge {\bf e}_{z} 
\wedge {\bf e}_{t}\:\rbrace \enspace, \nonumber\\ 
&&\quad i = 1, ... , 16 \enspace.
  \label{eq:02}
\end{eqnarray} 

\noindent
The covariant event, $x_\mu = g_{\mu\nu}\:x^\nu$, can be obtained 
with the help of the metric tensor $g_{\mu\nu}$ that we assume here to 
be diagonalized, i.e., 
$g_{\mu\nu} = \mathrm{diag}(g_{11}, g_{22}, g_{33}, g_{44})$.
E.g., with $g_{11} = g_{22} = g_{33} = -1$  and $g_{44} = +1$, 
it follows that $x_\mu = (-x, -y, -z, t)$.

\noindent
For the four given 
events $x^\mu_{1} = (x_{1}, y_{1}, z_{1}, t_{1})$, 
$x^\mu_{2} = (x_{2}, y_{2}, z_{2}, t_{2})$, 
$x^\mu_{3} = (x_{3}, y_{3}, z_{3}, t_{3})$, 
and $x^\mu_{4} = (x_{4}, y_{4}, z_{4}, t_{4})$, we form the
differences
\begin{eqnarray}
&&dx_{1}\:=\:x_{2} - x_{1} \enspace, \enspace
  dx_{2}\:=\:x_{3} - x_{1} \enspace, \enspace
  dx_{3}\:=\:x_{4} - x_{1} \enspace, \nonumber\\
&&dy_{1}\:=\:y_{2} - y_{1} \enspace, \enspace
  dy_{2}\:=\:y_{3} - y_{1} \enspace, \enspace
  dy_{3}\:=\:y_{4} - y_{1} \enspace, \nonumber\\
&&dz_{1}\:=\:z_{2} - z_{1} \enspace, \enspace
  dz_{2}\:=\:z_{3} - z_{1} \enspace, \enspace
  dz_{3}\:=\:z_{4} - z_{1} \enspace, \nonumber\\
&&dt_{1}\:=\:t_{2} - t_{1} \enspace, \enspace
  dt_{2}\:=\:t_{3} - t_{1} \enspace, \enspace
  dt_{3}\:=\:t_{4} - t_{1} \enspace.
  \label{eq:03}
\end{eqnarray}

\noindent
Let us define the determinants
\begin{eqnarray}
dS^{ikl}\:=\:\left|
\begin{array}{ccc}
dx_{1}^{i} & dx_{2}^{i} & dx_{3}^{i}\\
\vspace*{-0.25cm}\\
dx_{1}^{k} & dx_{2}^{k} & dx_{3}^{k}\\
\vspace*{-0.25cm}\\
dx_{1}^{l} & dx_{2}^{l} & dx_{3}^{l}
\end{array}
\right| \enspace , \enspace i,k,l \in \{1,2,3,4\}
\end{eqnarray}

\noindent
where $dx_{m}^{1} = dx_{m}$, $dx_{m}^{2} = dy_{m}$, $dx_{m}^{3} = dz_{m}$, 
$dx_{m}^{4} = dt_{m}$, $m = 1, 2, 3$. 
\\
\indent
E.g., while using Geometric Algebra (\textit{cf.}, 
e.g., Ref.~\cite{PERW09}) one can show that the contravariant $4$-normal 
is given through 
\begin{equation}
	d\sigma^\mu = \frac{1}{6}\left(\frac{dS^{234}}{g_{11}},
                      \frac{dS^{143}}{g_{22}},\frac{dS^{124}}{g_{33}},
                      \frac{dS^{132}}{g_{44}}\right) \enspace .
\end{equation}

\noindent
Note that the corresponding covariant $4$-normal,
\begin{equation}
	d\sigma_\mu = g_{\mu\nu}\:d\sigma^\nu 
        = \textstyle{\frac{1}{6}}\:(dS^{234}, dS^{143}, dS^{124}, dS^{132}) \enspace ,
\end{equation}

\noindent
does not depend on the $4$D metric under consideration.
For the special case of an isochronous hypersurface, i.e., 
$dt_{1} = dt_{2} = dt_{3} = 0$ (\textit{cf.},
subsection~\ref{subsec:isochronous}, and Eq.~(\ref{eq:03})), we get
\begin{equation}
	d\sigma_\mu = \textstyle{\frac{1}{6}}\:(0, 0, 0, dS^{132}) \enspace ,
\end{equation}

\noindent
where $dS^{132} = -\,dS^{123}$, i.e., $dS^{132}$ equals the 
\textit{negative} volume 
of a parallelepiped that is spanned by the contravariant $4$-vectors 
$dx^\mu_{m} = (dx_{m}, dy_{m}, dz_{m}, 0)$, $m = 1, 2, 3$.
Eventually, while not considering relativity, i.e.,
$g_{\mu\nu} = \mathrm{diag}(+1, +1, +1, +1)$, both $d\sigma^\mu$ and
$d\sigma_\mu$ become identical, i.e., 
$d\sigma^\mu \equiv d\sigma_\mu$.
\\
\indent
The previous evaluations have led to the particular
$\oplus$/$\ominus$ naming convention of 
Table~\ref{tab:02} used in $4$D.
The chosen orientations therein, together with the particular choices 
of Eqs.~(\ref{eq:02}) and~(\ref{eq:03}) yield a $4$-normal vector 
with all of its components larger than zero for the oriented tetrahedron 
shown in Fig.~\ref{fig:27}.c.

\section{Subspace Orientations}
\label{app:appB}

\begin{figure}[t]
  \begin{center}
    \scalebox{0.58}{
    \epsfig{width=10.0cm,figure=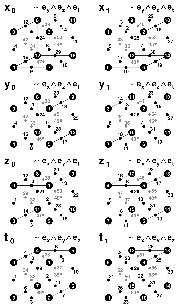}
    } 
  \end{center}
  \vspace{-0.3cm} 
  \caption{
$3$D subspaces for the $4$D-cell shown in Fig.~\ref{fig:07}.
Note that the indices for toxels (white), 
boundary cube centers (black), and junctures (gray), coincide 
with the indexing as shown in Fig.s~\ref{fig:07}.a -- \ref{fig:07}.c.
The $3$-blades are elements of the canonical multilinear basis, 
Eq.~(\ref{eq:02}) (see text).
} 
  \label{fig:28}
\end{figure}

\noindent
A tesseract is bounded by eight cubes.
In Fig.~\ref{fig:28}, we picture the eight bounding cubes
(or $3$D subspaces) of the tesseract shown (in NV) in 
Fig.~\ref{fig:07} at fixed components, $x_0$, $x_1$, $y_0$, $y_1$, 
$z_0$, $z_1$, $t_0$, and $t_1$.
The indices $0$ and $1$ refer to the minimum and maximum bounds of the
corresponding dimensions, respectively.
Each $3$D subspace is labeled here with toxel site
IDs (\textit{cf.}, Fig.~\ref{fig:07}.a), boundary cube centers
(\textit{cf.}, Fig.~\ref{fig:07}.b), and junctures (\textit{cf.},
Fig.~\ref{fig:07}.c).
Note that the subspaces at $x_{0/1}$, $y_{0/1}$, $z_{0/1}$, and $t_{0/1}$,
are directly proportional to the basis $3$-blades ${\bf E}_{15}$, 
${\bf E}_{14}$, ${\bf E}_{13}$, and ${\bf E}_{12}$, respectively 
(\textit{cf.}, Fig.~\ref{fig:28} and Eq.~(\ref{eq:02})).
\\
\indent
If one applies the marching variant of VESTA~\cite{BRS12} to each 
one of these eight subspaces, one can emulate the processing steps 
of the STEVE algorithm as shown in Fig.s~\ref{fig:03}.b 
through~\ref{fig:03}.g.
However, because of
\begin{eqnarray}
{\bf e}_{x} \wedge {\bf e}_{y} \wedge {\bf e}_{z} \wedge {\bf e}_{t} 
=\:{\bf e}_{x} \wedge {\bf E}_{15} \:=\:{\bf e}_{y} \wedge (-{\bf E}_{14})&& \nonumber \\
=\:{\bf e}_{z} \wedge {\bf E}_{13} \:=\:{\bf e}_{t} \wedge (-{\bf E}_{12})&,&
\end{eqnarray}

\noindent
one then has \textit{by all means} to choose inverse 
orientations for the generated VESTA $N$-cycles in the $x_0$-, $y_1$-, 
$z_0$-, and $t_1$-subspaces.
Only this warrants the consistent, orientation preserving construction 
of the higher-dimensional polytopes (\textit{cf.}, Fig.~\ref{fig:03}.h).

\bibliographystyle{plain}

\end{document}